\documentclass[preprint, 12pt]{elsarticle}
\usepackage{amssymb}
\usepackage{amsmath,bm}
\usepackage{amsthm}
\usepackage{empheq}
\usepackage{color}
\usepackage{cancel}
\usepackage[scale=0.8]{geometry}
\usepackage{mathabx}
\usepackage[dvipsnames]{xcolor}

\title{Periodic Shadowing Sensitivity Analysis of Chaotic Systems}

\author[soton]{Davide Lasagna}\ead{davide.lasagna@soton.ac.uk}
\author[soton]{Ati Sharma}\ead{a.sharma@soton.ac.uk}
\author[soton,kuleuven]{Johan Meyers\fnref{fn1}}\ead{johan.meyers@kuleuven.be}

\address[soton]{Faculty of Engineering and the Environment, University of Southampton, Highfield, Southampton SO17 1BJ, UK}

\address[kuleuven]{KU Leuven, Mechanical Engineering, Celestijnenlaan 300A, B3001 Leuven, Belgium}

\fntext[fn1]{Visiting professor at the University of Southampton from September 2017 until February 2018, as part of a sabbatical leave, sponsored by the Flemish Science Foundation. }
 
\begin{document}

\begin{abstract}
The sensitivity of long-time averages of a hyperbolic chaotic system to parameter perturbations can be determined using the shadowing direction, the uniformly-bounded-in-time solution of the sensitivity equations. Although its existence is formally guaranteed for certain systems, methods to determine it are hardly available. One practical approach is the Least-Squares Shadowing (LSS) algorithm (Q Wang, SIAM J Numer Anal 52, 156, 2014), whereby the shadowing direction is approximated by the solution of the sensitivity equations with the least square average norm. Here, we present an alternative, potentially simpler shadowing-based algorithm, termed periodic shadowing. The key idea is to obtain a bounded solution of the sensitivity equations by complementing it with periodic boundary conditions in time. We show that this is not only justifiable when the reference trajectory is itself periodic, but also possible and effective for chaotic trajectories. Our error analysis shows that periodic shadowing has the same convergence rates as LSS when the time span $T$ is increased: the sensitivity error first decays as $1/T$ and then, asymptotically as $1/\sqrt{T}$. We demonstrate the approach on the Lorenz equations, and also show that, as $T$ tends to infinity, periodic shadowing sensitivities converge to the same value obtained from long unstable periodic orbits (D Lasagna, SIAM J Appl Dyn Syst 17, 1, 2018) for which there is no shadowing error. Finally, finite-difference approximations of the sensitivity are also examined, and we show that subtle non-hyperbolicity features of the Lorenz system introduce a small, yet systematic, bias. 
\end{abstract}
\maketitle

\section{Introduction}
Simulation-based modelling of dynamical systems has become a key element across many applications in engineering and physical sciences. For system analysis and design, the aim is typically to understand how certain quantities of interest depend on a set of design variables parametrising the system at hand. Linear sensitivity analysis methods are used for this purpose, often in the form of an adjoint method \citep{Cacuci:1981jk,Mohammadi:2004dg,Luchini:2014fv}.

For unsteady dynamical systems, techniques from optimal control theory are used \citep{Bewley:2001gx}, whereby the linearised equations are marched in time to examine the effect of small parameter perturbations on the {future} evolution of the system, starting from the same given initial condition. However, chaotic systems with unstable dynamics display a high sensitivity to parameter perturbations. Hence, exponentially growing modes feature prominently into the solution of the linearised equations \citep{LEA:2000cr}. When the time span is increased, with the aim of obtaining the sensitivity of converged long-time statistics, the sensitivity does not converge but rather grows exponentially in time, resulting in unphysical gradients.

Several remedies have been proposed, ranging from ensemble-average approaches \citep{Bewley:2001gx, Eyink:2004gk}, to methods based on the analysis of the invariant probability density function and its adjoint \citep{Thuburn:2005ez,Blonigan:2014kp}. Both approaches are, however, affected by severe computational issues, namely the slow sub-central-limit-theorem convergence of the ensemble-average approach and the explosive growth of the computational cost with the increase of the attractor dimension for the adjoint density approach, respectively.

A major advance has been obtained recently \citep{Wang:2013cx} by exploiting the so-called Shadowing Lemma, an established theoretical result in dynamical systems theory due to Bowen \citep{Bowen:1975hc}, that exclusively applies to systems with hyperbolic dynamics. This Lemma is better known in the computational sciences community for its use in justifying finite-precision calculations of chaotic trajectories affected by round-off error. In such a context, it asserts that there exist an exact trajectory that starts from a slightly different initial condition and remains uniformly close to (it shadows) the numerically generated ``noisy'' trajectory \citep{Hammel:1987bv}.
\begin{figure}[t]
	\centering
	\includegraphics[width=0.65\textwidth]{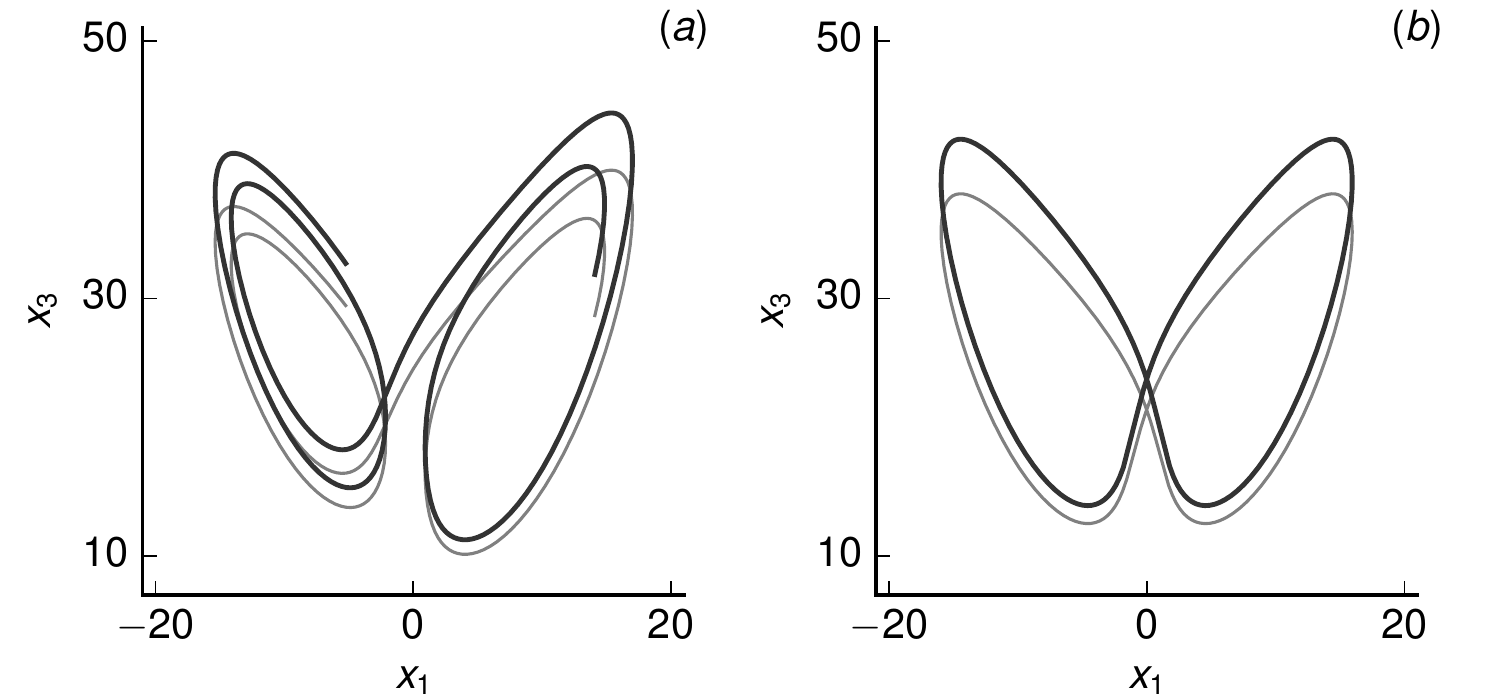}
	\caption{Shadowing for chaotic and periodic trajectories of the Lorenz equations for the perturbation parameter $\gamma$, defined in section \ref{sec:numerical-example}, equation (\ref{eq:lorenz-equations}). The shadow trajectory (\textcolor[rgb]{0.5, 0.5, 0.5}{\raisebox{0.07cm}{\rule{0.4cm}{0.2mm}}}), a solution of the perturbed equations at $\gamma=0.9$ stays uniformly close in time to the reference trajectory ({\raisebox{0.07cm}{\rule{0.4cm}{0.3mm}}}), a solution of the unperturbed equations at $\gamma=10$. }
	\label{fig:invariant-sets}
\end{figure}
In the context of sensitivity analysis of dynamical systems, the Shadowing Lemma can be used to show the existence of a trajectory of the perturbed system that starts at a different initial condition and remains uniformly close in time to the trajectory of the unperturbed system \citep{Wang:2013cx}. The concept is illustrated in figure \ref{fig:invariant-sets}-(a), for the Lorenz equations \citep{Lorenz:1963tf} defined in section \ref{sec:numerical-example}. Since the two trajectories remain uniformly close to each other, the linearisation holds throughout and accurate gradients can be obtained.

Although the Shadowing Lemma guarantees the existence of the shadowing direction, it does not suggest practical algorithms to determine it. One special case is that of periodic trajectories, depicted in figure \ref{fig:invariant-sets}-(b), where the shadowing direction is periodic in time \citep{Ruelle:1999bm}. Here, the topology of the problem can be introduced to derive periodic boundary conditions in time for the sensitivity equations, as recently shown in Ref.~\citep{Lasagna:2017tz}.  For chaotic trajectories, however, it currently appears unlikely that an efficient strategy exists that can be used to provide exact initial/boundary conditions and ones needs to rely  on approximations. Wang \citep{Wang:2013cx} suggested to exploit the exponential dichotomy of the linear dynamics and solved the sensitivity equations forward/backward along the stable/unstable directions. The approach, however, requires knowledge of the full decomposition of the tangent space in stable/unstable directions, a computationally expensive task \citep{Ginelli:2007ja}. In subsequent work, the same author proposed a method known as Least Squares Shadowing (LSS) that does not require such knowledge \citep{Wang:2014hu}. LSS approximates the unknown shadowing direction by determining the solution of the sensitivity equations with the least square average norm over the time span $T$. The minimisation ensures that exponentially growing modes that would highly contribute to the solution norm are effectively controlled, so that the optimal solution remains bounded, thus providing useful gradients. Variations of the method suitable for high-dimensional systems, using multiple-shooting strategies, have been also recently presented \citep{Blonigan:2018gd}. 

The original contribution of this paper is a novel shadowing-based algorithm, based on an alternative heuristic to approximate the shadowing direction. The key idea of the present method is to enforce periodic boundary condition in time to the sensitivity equations, leading to the name periodic shadowing. Providing such boundary conditions directly not only results in a method that is potentially simpler, but it sufficient to obtain bounded (periodic) solutions almost always, resulting in accurate gradients. The paper includes a detailed error analysis section, where we shown that the proposed method has the same asymptotic convergence rates of LSS \citep{Wang:2014bt,Chater:2017dw}, and where we derive asymptotic statistical distribution of the sensitivity error.

The paper is structured as follows. In sections \ref{sec:forward-sensitivity-method} and \ref{sec:adjoint-method} the tangent and adjoint periodic shadowing methods are derived, respectively. More technical details, e.g.~on numerical methods, are left to the appendices. In section \ref{app:error-analysis}, a detailed error analysis of the method is presented. In section \ref{sec:numerical-example} we report a demonstration of the method on the Lorenz equations \citep{Lorenz:1963tf}. The main objective is to provide numerical evidence to support the theoretical considerations of section \ref{app:error-analysis} on a well-studied problem that has been considered in many previous studies on sensitivity of chaotic systems. Finally, in section \ref{eq:conclusions} conclusions are outlined and few outstanding issues for future analysis are listed.
 
\section{Periodic shadowing: tangent sensitivity method}\label{sec:forward-sensitivity-method}
Let us consider the autonomous dynamical system given by the evolution equation
\begin{equation}\label{eq:system}
\dot{\mathbf{x}}(t) = \mathbf{f}(\mathbf{x}(t), p),
\end{equation}
where $t$ is time, the dot denotes differentiation with respect to time, and $\mathbf{x}(t) \in \mathbb{R}^N$. On the right hand side of (\ref{eq:system}), $\mathbf{f}(\mathbf{x}(t), p) : \mathbb{R}^N\times\mathbb{R} \rightarrow \mathbb{R}^N$ is a vector function of $\mathbf{x}(t)$ that depends, additionally, on a scalar parameter $p$ (this can be easily generalised to situations with more parameters). We assume that this vector function is sufficiently smooth with respect to its arguments, so that existence and uniqueness of solutions is formally guaranteed. When clear from the context, we also use the shorter notation $\mathbf{f}(t)$. Trajectories of (\ref{eq:system}) originating at some point $\mathbf{x}_0$ depend on the parameter too and are denoted as $\mathbf{x}(t; \mathbf{x}_0, p)$, i.e.~$\mathbf{x}(0; \mathbf{x}_0, p) = \mathbf{x}_0$. We drop the explicit dependence on the trailing arguments if they are clear from the context. 

Let now $J(\mathbf{x}(t), p) : \mathbb{R}^N\times\mathbb{R} \rightarrow \mathbb{R}$ be a scalar-valued functional, an observable of interest. Its finite-time average, denoted as
\begin{equation}\label{eq:time-average}
	\mathcal{J}^T(\mathbf{x}_0, p) = \frac{1}{T}\int_0^T J(\mathbf{x}(t; \mathbf{x}_0, p), p)\;\mathrm{d}t,
\end{equation}
will generally depend on the initial condition and the parameter value $p$. However, assuming ergodicity, the infinite-time average
\begin{equation}\label{eq:long-time-average}
	\mathcal{J}^\infty(p) = \lim_{T\rightarrow\infty} \mathcal{J}^T(\mathbf{x}_0, p)
\end{equation}
will not depend on the initial condition $\mathbf{x}_0$, but only on the parameter $p$. Understanding how the infinite-time average changes with $p$ is of paramount importance in many applications. At first order, for small perturbations around some reference $p$, this information is encoded by the gradient $\mathcal{J}^\infty_{\mathrm{d}p}(p)$, defined by the limit
\begin{align}\label{eq:derivative-as-limit}
	\displaystyle\mathcal{J}^\infty_{\mathrm{d}p}(p) =& \lim_{\delta p \rightarrow 0}\frac{1}{\delta p}\Big[\mathcal{J}^{\infty}(p^\prime) - \mathcal{J}^\infty(p)\Big]\nonumber \\ =& \lim_{\delta p \rightarrow 0} \frac{1}{\delta p}\Big[\lim_{T^\prime\rightarrow\infty}\mathcal{J}^{T^\prime}(\mathbf{x}^\prime_0, p^\prime) - \lim_{T\rightarrow\infty}\mathcal{J}^T(\mathbf{x}_0, p) \Big],
\end{align}
where $\delta p = p^\prime - p$ is the parameter perturbation and $\mathbf{x}_0$ and $\mathbf{x}^\prime_0$ are arbitrary initial conditions because of the ergodicity assumption. The existence of this limit, i.e.~the differentiability of the infinite-time averages of a dynamical system, is a long-standing question in dynamical systems theory, but, for instance, it can be shown to exists for uniformly hyperbolic systems \citep{Ruelle:1997et, Ruelle:2009eo}. At this stage we assume such limit exists. 

In (\ref{eq:derivative-as-limit}), the point $\mathbf{x}^\prime_0$ is the origin of the trajectory $\mathbf{x}^\prime(t; \mathbf{x}^\prime_0, p^\prime)$, satisfying the perturbed system
\begin{equation}\label{eq:system-perturbed}
	\dot{\mathbf{x}}^\prime(t) = \mathbf{f}(\mathbf{x}^\prime(t), p^\prime)
\end{equation}
over a time span $[0, T^\prime]$, where $T^\prime = T + \delta T$ can be arbitrarily selected since it does not affect the $T^\prime\rightarrow\infty$ limit in (\ref{eq:derivative-as-limit}). To obtain the gradient (\ref{eq:derivative-as-limit}) using a linear method, we first define the difference between the perturbed and reference trajectories as
\begin{equation}\label{eq:correct-difference}
	\delta \mathbf{x}(t) = \mathbf{x}^\prime(t{T^\prime}/{T}) - \mathbf{x}(t),
\end{equation}
such that the difference is defined over $[0, T]$, but time actually spans the full interval $[0, T^\prime]$ on the perturbed trajectory. In other words, $t\in [0, T]$ is now the independent variable parametrising trajectories of the perturbed system. In the literature of periodic systems, this approach is known as the Linstedt-Poincar\`e technique \citep{Viswanath:2001ko}. If the same time span was used, the difference (\ref{eq:correct-difference}) would not be periodic with period $T$, but would contain algebraically growing modes. 

We now assume, and will later verify, that the difference (\ref{eq:correct-difference}) remains small for well-selected conditions $\mathbf{x}^\prime_0$, such that the linearisation
\begin{equation}\label{eq:expand-difference}
	\delta \mathbf{x}(t) = \mathbf{y}(t)\delta p + \mathcal{O}(\delta p^2)
\end{equation}
holds throughout. In what follows, the quantity $\mathbf{y}(t)$ will be referred to as the sensitivity. The sensitivity is then used to linearise the observable around the reference trajectory as
\begin{equation}\label{eq:expand-observable}
	J(\mathbf{x}^\prime(tT^\prime/T), p^\prime) = J(\mathbf{x}(t), p) + J_{\partial \mathbf{x}}(\mathbf{x}(t), p)\cdot\mathbf{y}(t)\delta p + J_{\partial p}(\mathbf{x}(t), p)\delta p + \mathcal{O}(\delta p^2), \quad t \in [0, T], 
\end{equation}
so that the limit (\ref{eq:derivative-as-limit}) can be expressed as
\begin{align}\label{eq:sensitivity-limit-1}
	\mathcal{J}^\infty_{\mathrm{d}p}(p) =& \lim_{\delta p \rightarrow 0} \  \lim_{T, T^\prime\rightarrow\infty} \frac{1}{\delta p}\Bigg[\frac{1}{T^\prime}\int_0^{T^\prime} J(\mathbf{x}^\prime(t; \mathbf{x}_0^\prime, p^\prime), p^\prime)\;\mathrm{d}t - \frac{1}{T}\int_0^T J(\mathbf{x}(t; \mathbf{x}_0, p), p)\;\mathrm{d}t\Bigg]\nonumber\\
	=& \lim_{\delta p \rightarrow 0} \ \lim_{T\rightarrow\infty}\frac{1}{\delta p}\Bigg[\frac{1}{T^\prime} \int_0^{T} J(\mathbf{x}^\prime(t\frac{T^\prime}{T}; \mathbf{x}_0^\prime, p^\prime), p^\prime)\;\mathrm{d}t\frac{T^\prime}{T} - \frac{1}{T}\int_0^T J(\mathbf{x}(t; \mathbf{x}_0, p), p)\;\mathrm{d}t\Bigg] \nonumber \\
	=& \lim_{T\rightarrow\infty} \frac{1}{T}\int_0^{T} J_{\partial p}(\mathbf{x}(t), p) +  J_{\partial \mathbf{x}}(\mathbf{x}(t), p)\cdot\mathbf{y}(t)\;\mathrm{d}t.
\end{align}
In (\ref{eq:sensitivity-limit-1}), the upper limit of integration of the first integral in the second step has been changed from $T^\prime$ to $T$ using the time rescaling $t^\prime = tT^\prime/T$ implicitly defined by (\ref{eq:correct-difference}), where $t^\prime \in [0, T^\prime]$. The evolution equation for $\mathbf{y}(t)$ is derived by differentiating (\ref{eq:correct-difference}) with respect to $t$, obtaining
\begin{equation}
	\dot{\delta\mathbf{x}}(t) = \frac{T^\prime}{T}\mathbf{f}(\mathbf{x}^\prime(t T^\prime/T), p^\prime) - \mathbf{f}(\mathbf{x}(t), p),
\end{equation}
where the factor $T^\prime/T$ arises because points $\mathbf{x}^\prime(tT^\prime/T)$ on the perturbed trajectory move at a different rate than usual when $t$ varies. Linearising the vector field around $\mathbf{x}(t)$, noting that to first order $T^\prime/T = 1 + T_{\mathrm{d} p}/T\delta p$, dividing by $\delta p$ and taking the limit for $\delta p \rightarrow 0$ leads to the sensitivity equations
\begin{equation}\label{eq:sensitivity-equation}
	\dot{\mathbf{y}}(t) = \mathbf{f}_{\partial \mathbf{x}}(\mathbf{x}(t), p)\cdot\mathbf{y}(t) + \mathbf{f}_{\partial p}(\mathbf{x}(t), p) + \frac{T_{\mathrm{d} p}}{T}\mathbf{f}(\mathbf{x}(t), p),
\end{equation}
where $\mathbf{f}_{\partial \mathbf{x}}(\mathbf{x}(t), p) \in \mathbb{R}^{N\times N}$ is the system Jacobian containing the partial derivatives of the vector field with respect to the state space coordinates whilst $\mathbf{f}_{\partial p}(\mathbf{x}(t), p) \in \mathbb{R}^{N}$ is a vector containing the partial derivatives of the vector field with respect to the parameter. Note that the gradient $T_{\mathrm{d} p}$ is still an unknown and arbitrary quantity, because we have not yet specified how the time spans $T$ and $T^\prime$ should be related when the $\delta p\rightarrow0$ limit is taken.

Stepping back to the limit (\ref{eq:derivative-as-limit}), we observe that the initial condition $\mathbf{x}^\prime_0$ can be selected arbitrarily because of the ergodicity assumption. In a linearised setting, this corresponds to selecting an arbitrary $\mathbf{y}(0)$. Classical sensitivity analysis methods select the initial condition $\mathbf{y}(0) = \mathbf{0}$, the linearisation of $\mathbf{x}^\prime_0 = \mathbf{x}_0$. In the adjoint method, this results in an homogeneous terminal condition in the adjoint problem. This choice arises from optimal control theory ideas, where the focus is typically on the effects of parameter perturbations on the {future} evolution of the system, starting from the same initial condition. However, it is well known that two trajectories originating at the same point separate initially at an exponential rate, and the difference saturates in a finite-time around a finite value due to global boundedness. The linearised equations (\ref{eq:sensitivity-equation}), however, do not model these nonlinear effects, and thus $\mathbf{y}(t)$ continues growing at an average exponential rate for all $t$ \citep{LEA:2000cr}. In other words, for any finite $\delta p$, there is a finite $T$ at which the linearisation fails and higher order terms neglected in (\ref{eq:expand-difference}) and (\ref{eq:expand-observable}) become important \citep{Thuburn:2005ez}. This growth is reflected in an unphysical exponential increase of the gradient (\ref{eq:sensitivity-limit-1}) as $T$ is increased \citep{LEA:2000cr,Eyink:2004gk}.

\begin{figure}[t]
	\centering
	\includegraphics[width=0.65\textwidth]{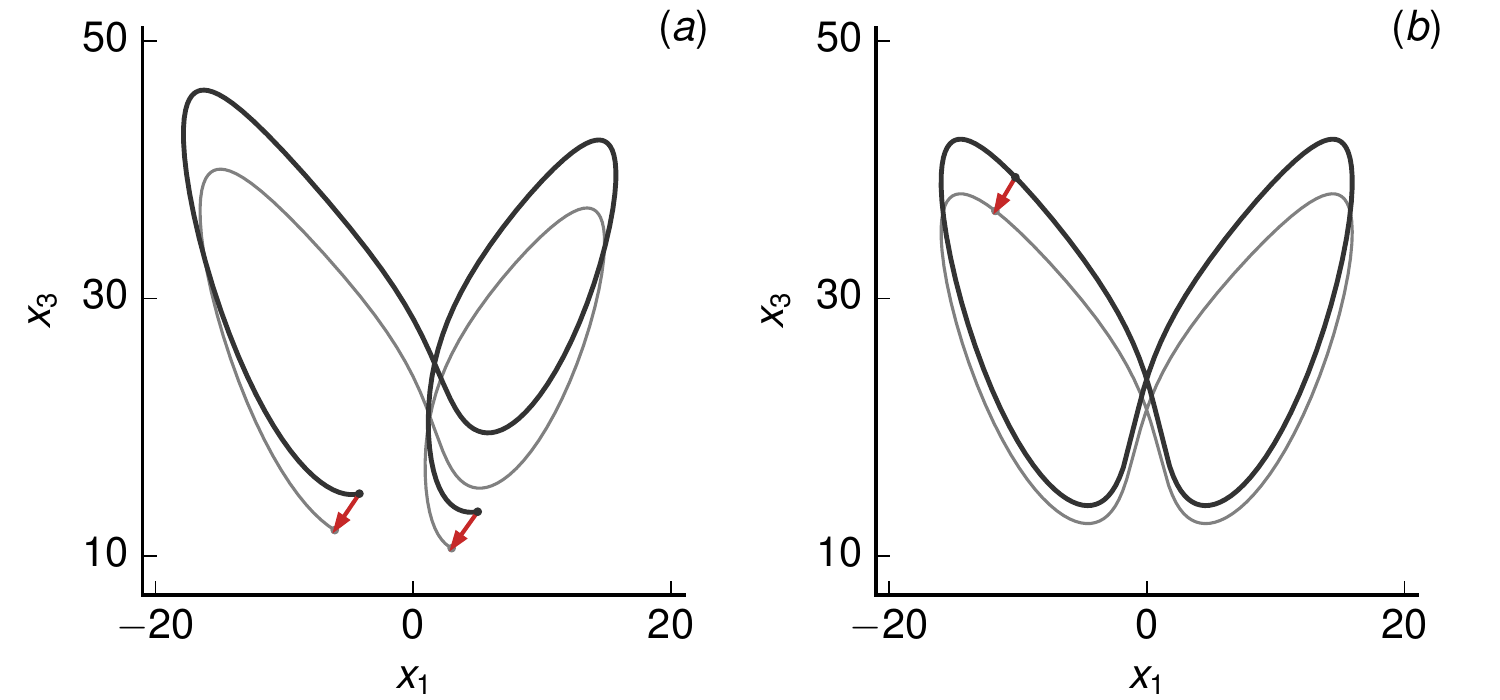} 
	\caption{Geometry of the periodicity condition (\ref{eq:boundary-conditions-key-idea}), for aperiodic (a) and periodic (b) trajectories of the Lorenz equations with $\gamma$ as the perturbation parameter (see \S\ref{sec:numerical-example} for details, same colours as in figure \ref{fig:invariant-sets}).}
	\label{fig:same-direction}
\end{figure}

In order for the linearisation to remain valid, and thus for the gradient (\ref{eq:sensitivity-limit-1}) to converge as $T\rightarrow\infty$, the sensitivity $\mathbf{y}(t)$ should remain bounded. As discussed in the introduction, this is indeed not just possible, but also theoretically justified by the Shadowing Lemma \citep{Wang:2014hu} for certain classes of systems. However, the same Lemma does not specify algorithms to determine it in practice, e.g. it does not provide initial or boundary conditions that can be used to solve (\ref{eq:sensitivity-equation}). The original contribution of this paper is that we propose such conditions. In the nonlinear setting, the key idea is illustrated in panel (a) of figure \ref{fig:same-direction}, for a chaotic trajectory of the Lorenz equations (see \S\ref{sec:numerical-example} for details). We impose the condition that the end points of the perturbed trajectory move in the same unspecified direction by the same unspecified amount, indicated by the arrows. Mathematically, this is expressed by the boundary conditions
\begin{equation}\label{eq:boundary-conditions-key-idea}
	\mathbf{x}_0 - \mathbf{x}^\prime_0 = \mathbf{x}(T; \mathbf{x}_0, p) - \mathbf{x}^\prime(T^\prime; \mathbf{x}_0^\prime, p^\prime).
\end{equation}
The linearisation of (\ref{eq:boundary-conditions-key-idea}), obtained by dividing both sides by $\delta p$ and taking the $\delta p \rightarrow 0$ limit, leads to the periodic boundary conditions
\begin{equation}\label{eq:linear-boundary-conditions}
	\mathbf{y}(0) = \mathbf{y}(T),
\end{equation}
hence the name periodic shadowing.
These conditions are not sufficient to determine the gradient $T_{\mathrm{d} p}$, which remains arbitrary. In fact, the solution of the sensitivity equations will vary when $T_{\mathrm{d} p}$ is arbitrarily changed and so will the gradient (\ref{eq:sensitivity-limit-1}). A strategy to select a value of $T_{\mathrm{d} p}$ and identify a specific solution is therefore required. Forcibly setting $T_{\mathrm{d} p} = 0$ does not lead to an accurate method, essentially because neglecting the growth of algebraic modes produces a spurious sensitivity error that does not vanish as $T\rightarrow\infty$. Here, we propose to use an approach that is frequently employed in bifurcation analysis and continuation for periodic systems and is straightforward to use in many numerical methods. Specifically, rather than fixing the gradient $T_{\mathrm{d} p}$ a priori, we impose that the solution of (\ref{eq:sensitivity-equation}) satisfies the additional orthogonality condition 
\begin{equation}\label{eq:phase-locking-constraint}
	\mathbf{f}(\mathbf{x}(0), p)^\top\cdot\mathbf{y}(0) = 0.
\end{equation}
This constraint implicitly fixes $T_{\mathrm{d} p}$ to a value that we denote in what follows as $T_{\mathrm{d} p}^0$. We will show in section~\ref{app:error-analysis} that this approach 
leads to converging sensitivities for $T\rightarrow \infty$, and yields physically meaningful values of $T_{\mathrm{d} p}^0/T$ that have direct connection to the variation of the system's time scale under parameter perturbations.

Finally, combining the sensitivity equations (\ref{eq:sensitivity-equation}), the boundary conditions (\ref{eq:linear-boundary-conditions}) and the orthogonality constraint (\ref{eq:phase-locking-constraint}) leads to
\begin{subequations}\label{eq:forward-sensitivity-problem}
\begin{empheq}[left={\empheqlbrace}]{alignat=1}
\dot{\mathbf{y}}(t) =& \;\mathbf{f}_{\partial \mathbf{x}}(\mathbf{x}(t), p)\cdot\mathbf{y}(t) + \mathbf{f}_{\partial p}(\mathbf{x}(t), p) + \frac{T_{\mathrm{d} p}}{T}\mathbf{f}(\mathbf{x}(t), p),\quad t\in [0, T] \label{eq:forward-sensitivity-problem-a}\\
\mathbf{y}(0) =&\;  \mathbf{y}(T) \label{eq:forward-sensitivity-problem-b}\\
0=&\; \mathbf{f}(\mathbf{x}(0), p)^\top\cdot\mathbf{y}(0), \label{eq:forward-sensitivity-problem-c}
\end{empheq}
\end{subequations}
the tangent periodic shadowing problem. From a mathematical viewpoint, (\ref{eq:forward-sensitivity-problem}) is a boundary value problem (BVP). The periodic boundary conditions ensure that exponentially growing modes do not feature in the solution, regardless of $T$. However, as a consequence, a dedicated numerical method for boundary value problems is required. Fortunately, the structure of this BVP is similar to that arising in bifurcation and continuation problems of periodic orbits of dynamical systems \citep{Govaerts:2000vv}. This means that efficient numerical methods to solve such problems, applicable to systems of both small and very large dimension, are well developed. The major challenges stem primarily from the instability of (\ref{eq:forward-sensitivity-problem-a}) and, potentially, from the large dimensionality for discretisations of PDEs. For the numerical example discussed in this paper, the Lorenz equations, we have used a classical method based on multiple-shooting, where dense linear algebra methods have been used \citep{Ascher:1994ty}. A more detailed discussion of numerical methods is deferred to \ref{sec:multiple-shooting}.

Note that the structure of (\ref{eq:forward-sensitivity-problem}) is identical to that arising in the tangent sensitivity method recently reported in \citep{Lasagna:2017tz} for unstable periodic orbits (UPO) of chaotic dynamical systems. In the case of a periodic reference trajectory, $\mathbf{x}(T; \mathbf{x}_0, p) = \mathbf{x}_0$, there is no approximation involved in the choice of periodic boundary conditions (\ref{eq:linear-boundary-conditions}), because the shadowing direction is itself periodic. 
As illustrated in panel (b) of figure \ref{fig:same-direction}, the initial and final points of the trajectory move in the same direction when parameters are varied because they are precisely the same point. The difference with (\ref{eq:forward-sensitivity-problem}) is that the time dependent coefficients on the right hand side of (\ref{eq:forward-sensitivity-problem-a}) are not periodic on a chaotic trajectory, hence periodicity is not guaranteed for the derivatives of the solution $\mathbf{y}(t)$, while it is for UPOs.

\section{Periodic shadowing: adjoint sensitivity method}\label{sec:adjoint-method}
In situations where the sensitivity of one or a few observables with respect to many parameters is required, an adjoint method is preferable. To obtain the adjoint periodic shadowing method we employ a classical Lagrangian approach \citep{Cacuci:1981jk,Borzi:2012ik} and start by constructing the finite-time Lagrangian function
\begin{equation}\label{eq:lagrangian}
	\mathcal{L}^T(\mathbf{x}_0, p) \equiv \mathcal{J}^T(\mathbf{x}_0, p) + \frac{1}{T}\int_0^T {\boldsymbol \lambda}^\top(t)\cdot \big[\dot{\mathbf{x}}(t) - \mathbf{f}(\mathbf{x}(t), p)\big]\;\mathrm{d}t,
\end{equation}
$\mathcal{L}^T : \mathbb{R} \rightarrow \mathbb{R} $, by adjoining the governing equation (\ref{eq:system}) to the cost function, with the adjoint variables $\boldsymbol \lambda (t)\in{ }\mathbb{R}^N$. Since the governing equation is satisfied for all $p$ and for all $t\in[0, T]$ along the trajectory, $\mathcal{L}^T(\mathbf{x}_0, p) = \mathcal{J}^T(\mathbf{x}_0, p)$ for every $p$ and thus $\mathcal{L}^T_{\mathrm{d} p} = \mathcal{J}^T_{\mathrm{d} p}$.  The derivative of the finite-time Lagrangian with respect to the parameter is defined similarly to (\ref{eq:sensitivity-limit-1}), as the $\delta p \rightarrow 0$ limit of the difference quotient
\begin{equation}
	\mathcal{L}^T_{\mathrm{d} p}(\mathbf{x}_0, p) = \lim_{\delta p \rightarrow 0} \frac{1}{\delta p} \Big[ \mathcal{L}^{T^\prime}(\mathbf{x}_0^\prime, p^\prime) - \mathcal{L}^T(\mathbf{x}_0, p)\Big],
\end{equation}
where $\mathcal{L}^{T^\prime}(\mathbf{x}_0^\prime, p^\prime)$ and $\mathcal{L}^T(\mathbf{x}_0, p)$ are defined over the reference and perturbed trajectories, respectively. Tedious, yet straightforward algebraic manipulations that use the same approach as for (\ref{eq:sensitivity-limit-1}) and (\ref{eq:sensitivity-equation}) lead to 
\begin{align}
	\mathcal{L}^T_{\mathrm{d} p}(p) =& \frac{1}{T}\int_0^{T} \underbrace{\big[J_{\partial \mathbf{x}}^\top(t) - \dot{\boldsymbol \lambda}^\top(t) - {\boldsymbol \lambda}^\top(t) \cdot \mathbf{f}_{\partial\mathbf{x}}(t)\big]}_{A}\cdot \mathbf{y}(t) + J_{\partial p}(t) - {\boldsymbol \lambda}^\top(t)\cdot\bigg[\mathbf{f}_{\partial p}(t) - \frac{T_{\mathrm{d} p}}{T}\mathbf{f}(t)\bigg]\;\mathrm{d}t \nonumber\\
	+& \underbrace{\frac{1}{T}\big[{\boldsymbol \lambda}^\top(t)\cdot\mathbf{y}(t)\big]\bigg|_{0}^T}_{B}
\end{align}
where the term $B$ arises from integration by parts in time, and terms multiplying the sensitivity $\mathbf{y}(t)$ have been collected into the term $A$. The adjoint variables are then selected such that the terms $A$ and $B$ vanish identically, to avoid the explicit computation of $\mathbf{y}(t)$ for every parameter of interest. Requiring the term $A$ to vanish leads to an adjoint equation, while it is straightforward to see that requiring that $B = 0$, and using the periodic boundary conditions of the tangent problem (\ref{eq:linear-boundary-conditions}), is equivalent to imposing periodic boundary conditions in time on the adjoint solution. This leads to the adjoint periodic shadowing problem
\begin{subequations}\label{eq:adjoint-sensitivity-problem}
\begin{empheq}[left={\empheqlbrace}]{alignat=1}
\dot{\boldsymbol \lambda}(t) &= -\mathbf{f}_{\partial \mathbf{x}}^\top(\mathbf{x}(t), p)\cdot{\boldsymbol \lambda}(t) + J_{\partial \mathbf{x}}(\mathbf{x}(t), p), \quad t \in [0, T], \label{eq:adjoint-sensitivity-problem-a}\\
{\boldsymbol \lambda}(0) &={\boldsymbol \lambda}(T). \label{eq:adjoint-sensitivity-problem-b}
\end{empheq}
\end{subequations}

Similarly to (\ref{eq:forward-sensitivity-problem}), the periodic boundary conditions guarantee that the adjoint solution does not exhibit the typical exponential growth observed in the classical backward integration of the adjoint equation from the homogeneous terminal condition ${\boldsymbol \lambda}(T) = \mathbf{0}$. Unlike in (\ref{eq:forward-sensitivity-problem}), the gradient $T_{\mathrm{d} p}$  does not feature in the adjoint problem, whose solution is thus unique and does not require an additional constraint. This gradient, however, features in the integral that, upon solution of (\ref{eq:adjoint-sensitivity-problem}), provides the sensitivity of the time average
\begin{equation}\label{eq:sensitivity-integral-adjoint}
	\mathcal{J}^T_{\mathrm{d}p}(\mathbf{x}_0, p) = \mathcal{L}^T_{\mathrm{d}p}(\mathbf{x}_0, p) = \frac{1}{T}\int_0^T J_{\partial p}(\mathbf{x}(t), p) - {\boldsymbol \lambda}^\top(t)\cdot\bigg[ \mathbf{f}_{\partial p}(\mathbf{x}(t), p) + \frac{T_{\mathrm{d} p}^0}{T}\mathbf{f}(\mathbf{x}(t), p)\bigg]\,\mathrm{d}t.
\end{equation}
As discussed for the tangent method, the quantity $T_{\mathrm{d} p}$ is selected by requiring that the orthogonality constraint (\ref{eq:phase-locking-constraint}) holds. In the adjoint method, the sensitivity $\mathbf{y}(t)$ is never computed explicitly and an alternative approach is required to obtain $T_{\mathrm{d} p}^0$. The technique we used in this paper requires solving one additional adjoint problem, with the same structure of (\ref{eq:adjoint-sensitivity-problem-a}), but with a different forcing term. This has implications on the overall computational cost of the adjoint method, which, depending on the numerical method used will range from one to two times the cost of solving (\ref{eq:adjoint-sensitivity-problem}), because the cost of computations can be amortised by solving the two adjoint problems together. The technical details and discussion on computational costs are reported in \ref{app:adjoint-tp0}.

\section{Error analysis}\label{app:error-analysis}
This section presents an error analysis of the periodic shadowing sensitivity method. As for other shadowing-based sensitivity methods, we assume hyperbolicity \citep{Kuznetsov:2012er}. For this class of systems, the differentiability of statistical quantities, and the existence of the limit (\ref{eq:derivative-as-limit}), is an important known result in dynamical systems theory \citep{Ruelle:1997et, Ruelle:2009eo}. We further assume boundedness of trajectories of the system (\ref{eq:system}) and smoothness of the observable of interest $J(\mathbf{x}(t), p)$.

For hyperbolic systems, the Shadowing Lemma \citep{Bowen:1975hc} has been used in recent work \citep{Wang:2013cx} to guarantee the existence of the shadowing direction, the unique trajectory $\mathbf{y}_{\mathrm{S}}(t)$ satisfying
\begin{equation}\label{eq:shadowing-direction-equation}
	\dot{\mathbf{y}}_\mathrm{S}(t) = \mathbf{f}_{\partial \mathbf{x}}(\mathbf{x}(t), p)\cdot \mathbf{y}_\mathrm{S}(t) + \mathbf{f}_{\partial p}(\mathbf{x}(t), p) + \eta(t)\mathbf{f}(\mathbf{x}(t), p), \quad t \in [-\infty, \infty],
\end{equation}
that is uniformly bounded in time. More precisely, there exists a finite positive constant $B$ such that 
\begin{equation}\label{eq:bound-B}
	\|\mathbf{y}_\mathrm{S}(t)\| \leq B.
\end{equation}
Boundedness implies that both exponentially and algebraically growing modes do not materialise in $\mathbf{y}_{\mathrm{S}}(t)$. This is possible by an appropriate definition of the initial condition $\mathbf{y}_\mathrm{S}(0)$ at some arbitrary initial time, to factor out the exponential modes, and an appropriate definition of the scalar time transformation term $\eta(t)$, to take care of the algebraic modes. To explain the effect of the transformation on the algebraic modes can we introducing the fundamental matrix solution $\mathbf{Y}(t, \mathbf{x}_0)$ of (\ref{eq:shadowing-direction-equation}), obeying the initial value problem
\begin{equation} \label{eq:introductionY}
\hspace{2cm}\dot{\mathbf{Y}}(t, \mathbf{x}_0) = \mathbf{f}_{\partial \mathbf{x}}(\mathbf{x}(t), p)\cdot\mathbf{Y}(t, \mathbf{x}_0),\quad t\in[0, T],\quad \mathbf{Y}(0, \mathbf{x}_0) =\; \mathbf{I},
\end{equation}
with $\mathbf{I}$ the identity matrix of appropriate size. We then split the time transformation term into its infinite-time mean and the associated fluctuation
\begin{equation}\label{eq:transformation-decomposition}
	\eta(t) = \bar{\eta}^\infty + \tilde{\eta}^\infty(t),
\end{equation}
to derive the general solution of (\ref{eq:shadowing-direction-equation})
\begin{equation}\label{eq:general-solution-shadow}
	\mathbf{y}_\mathrm{S}(t) = \mathbf{Y}(t, \mathbf{x}_0)\cdot\bigg[\mathbf{y}_\mathrm{S}(0) + \int_{0}^{t} \mathbf{Y}^{-1}(s, \mathbf{x}_0)\cdot\mathbf{f}_{\partial p}(\mathbf{x}(s), p)\mathrm{d}s\bigg] + \mathbf{f}(\mathbf{x}(t), p)\bigg[t\bar{\eta}^\infty + \int_{0}^t\tilde{\eta}^\infty(s)\mathrm{d}s\bigg].
\end{equation}
The last term illustrates how the mean component $\bar{\eta}^\infty$ takes care of the linear growth of algebraic modes, while the zero-mean component only controls the local shift of $\mathbf{y}_{\mathrm{S}}(t)$ along the vector field $\mathbf{f}(t)$ and can be arbitrarily chosen, e.g., to ensure that 
\begin{equation}\label{eq:orthogonality-constraint-shadow}
	\mathbf{f}(\mathbf{x}(t), p)^\top\cdot\mathbf{y}_\mathrm{S}(t) = 0
\end{equation}
for all $t$, leading to the ``canonical'' shadowing direction, as defined in Ref.~\citep{Chater:2017dw}.

If the shadowing direction were known, the sensitivity for the finite-time trajectory $\mathbf{x}(t; \mathbf{x}_0, p)$ defined over the time span $t\in[0, T]$ could be calculated as
\begin{align}\label{eq:sensitivity-for-shadowing}
	\mathcal{J}^{T, \mathrm{S}}_{\mathrm{d}p}(\mathbf{x}_0, p) =& \frac{1}{T}\int_0^{T} J_{\partial p}(\mathbf{x}(t), p) +  J_{\partial \mathbf{x}}(\mathbf{x}(t), p)^\top\cdot\mathbf{y}_{\mathrm{S}}(t) + \eta(t)[J(\mathbf{x}(t)) - \mathcal{J}^T]\;\mathrm{d}t
\end{align}
where the additional superscript $\mathrm{S}$ hints at the fact that $\mathbf{y}_{\mathrm{S}}(t)$ is used for the calculation. In Ref.~\citep{Chater:2017dw} it is proven by exchange of limits that the finite-time sensitivity (\ref{eq:sensitivity-for-shadowing}) converges to the infinite-time sensitivity $\mathcal{J}^{\infty}_{\mathrm{d} p}$, defined by the limit (\ref{eq:derivative-as-limit}), as $T\rightarrow\infty$. For ergodic, mixing dynamical systems, the central limit theorem dictates the average rate of convergence. Specifically, for large enough $T$, the finite-time sensitivity will exhibit a random error
\begin{equation}\label{eq:random-error}
	\mathcal{E}^T_0(\mathbf{x}_0, p) = \mathcal{J}^{T, \mathrm{S}}_{\mathrm{d} p}(\mathbf{x}_0, p) - \mathcal{J}^{\infty}_{\mathrm{d} p} = C^T_0(\mathbf{x}_0, p)/\sqrt{T},
\end{equation}
where $C^T_0(\mathbf{x}_0, p)$ is a constant that is statistically distributed according to a certain probability density function (PDF) that is independent of $T$, but only depends on the dynamics (\ref{eq:system}) and the choice of the observable.

The solution of the periodic shadowing problem (\ref{eq:forward-sensitivity-problem}), denoted in this section as $\mathbf{y}_\mathrm{P}(t)$, is an approximation of the shadowing direction $\mathbf{y}_\mathrm{S}(t)$. The shadowing error, defined as
\begin{equation}\label{eq:error-definition}
	\mathbf{e}(t) = \mathbf{y}_\mathrm{P}(t) - \mathbf{y}_\mathrm{S}(t),\quad t\in [0, T],
\end{equation}
can be readily obtained by differentiating (\ref{eq:error-definition}) with respect to time and using the appropriate linearised equations, leading to the BVP
\begin{subequations}\label{eq:error-equations}
\begin{empheq}[left=\empheqlbrace]{alignat=2}
	\dot{\mathbf{e}}(t) =&\; \mathbf{f}_{\partial \mathbf{x}}(\mathbf{x}(t), p)\cdot \mathbf{e}(t) + \big[T^0_{\mathrm{d}p}/T - \eta(t)\big]\,\mathbf{f}(\mathbf{x}(t), p), \quad t \in [0, T]\label{eq:error-equations-a}\\
	\mathbf{e}(0) =&\; \mathbf{e}(T) + \mathbf{r} \label{eq:error-equations-b}\\
	0 =&\; \mathbf{e}(0)^\top\cdot\mathbf{f}(0),\label{eq:error-equations-c}
\end{empheq}
\end{subequations}
with $\mathbf{r} = \mathbf{y}_\mathrm{S}(T) - \mathbf{y}_\mathrm{S}(0)$. The orthogonality condition (\ref{eq:error-equations-c}) follows directly from (\ref{eq:forward-sensitivity-problem-c}) and (\ref{eq:orthogonality-constraint-shadow}). The general solution of (\ref{eq:error-equations}) is
\begin{equation}\label{eq:general-solution-error-equation}
	\mathbf{e}(t) = \mathbf{Y}(t, \mathbf{x}_0)\cdot\mathbf{e}(0) + \mathbf{f}(\mathbf{x}(t))\left[T^0_{\mathrm{d}p} - \int_0^t\eta(s)\,\mathrm{d}s\right].
\end{equation}

Resting on the formal convergence guarantees of $\mathcal{J}^{T, \mathrm{S}}_{\mathrm{d} p}(\mathbf{x}_0, p)$, our strategy to show convergence of the periodic shadowing method consists in analysing the sensitivity error
\begin{equation}\label{eq:sensitivity-error-1}
	\mathcal{E}^T_1(\mathbf{x}_0, p) = \mathcal{J}^{T, \mathrm{P}}_{\mathrm{d} p}(\mathbf{x}_0, p) - \mathcal{J}^{T, \mathrm{S}}_{\mathrm{d} p}(\mathbf{x}_0, p) = \frac{1}{T}\int_0^T J_{\partial \mathbf{x}}(\mathbf{x}(t), p)^\top\cdot\mathbf{e}(t) - \eta(t)[J(\mathbf{x}(t), p) - \mathcal{J}^T] \,\mathrm{d}t.
\end{equation}
where $\mathcal{J}^{T, \mathrm{P}}_{\mathrm{d} p}(\mathbf{x}_0, p)$ is the sensitivity computed using $\mathbf{y}_{\mathrm{P}}(t)$, the periodic solution of (\ref{eq:forward-sensitivity-problem}). The main result of this section will be that
\begin{equation}\label{eq:first-thing-to-show}
	\mathcal{E}^T_1(\mathbf{x}_0, p) = C_1^T(\mathbf{x}_0, p)/T,
\end{equation}
where $C_1^T(\mathbf{x}_0, p)$ is a constant that, similarly to $C_0^T(\mathbf{x}_0, p)$, is statistically distributed according to a certain probability density function that, asymptotically, does not depend on $T$. 

The rapid $1/T$ decay implies that for some sufficiently large $T$ the shadowing error $\mathcal{E}^T_1(\mathbf{x}_0, p)$ will be, on average, smaller than that of the random error $\mathcal{E}_0^T(\mathbf{x}_0, p)$, and the total error
\begin{equation}\label{eq:total-sensitivity-error}
	\mathcal{E}^T(\mathbf{x}_0, p) = \mathcal{E}^T_0(\mathbf{x}_0, p) + \mathcal{E}^T_1(\mathbf{x}_0, p) = \frac{C_0^T(\mathbf{x}_0, p)}{\sqrt{T}} + \frac{C_1^T(\mathbf{x}_0, p)}{T}
\end{equation}
will be mostly dominated by the first term and not by the details of the shadowing sensitivity algorithm. Note that when using a periodic trajectory of period $T$ for sensitivity analysis \citep{Lasagna:2017tz}, the error $\mathcal{E}^T_1(\mathbf{x}_0, p)$ is identically zero, since the shadowing direction is periodic and is found from the solution of the tangent problem (\ref{eq:forward-sensitivity-problem}). Only the random error $\mathcal{E}^T_0(\mathbf{x}_0, p)$ affects the sensitivity results. This fact will be used in the results section to provide further validation to the proposed method.

The technical development leading to equation (\ref{eq:first-thing-to-show}) consists of two steps. First, in section \ref{sec:error-analysis-step1}, we derive an expression for the sensitivity error $\mathcal{E}^T_1(\mathbf{x}_0, p)$ in terms of $\|\mathbf{e}(0)\|$, the norm of the shadowing error at the initial point. Then, we examine the behaviour of this term in section \ref{sec:showing-e0}.

\subsection{Obtaining the sensitivity error $\mathcal{E}^T_1(\mathbf{x}_0, p)$ in terms of $\|\mathbf{e}(0)\|$}\label{sec:error-analysis-step1}
To proceed, we consider individually the three sensitivity errors
\begin{equation}
	\mathcal{E}^T_{1}(\mathbf{x}_0, p) = \mathcal{E}^T_{1, -}(\mathbf{x}_0, p) + \mathcal{E}^T_{1, 0}(\mathbf{x}_0, p) + \mathcal{E}^T_{1, +}(\mathbf{x}_0, p),
\end{equation}
arising from the three components of the shadowing error,
\begin{equation}\label{eq:error-split}
\mathbf{e}(t) = \mathbf{e}^-(t) + \mathbf{e}^0(t) + \mathbf{e}^+(t),
\end{equation}
lying on the stable $V_{\mathbf{x}(t)}^-$, neutral $V_{\mathbf{x}(t)}^0$ and unstable $V_{\mathbf{x}(t)}^+$ linear subspaces at $\mathbf{x}(t)$, assumed to be disjoint due to hyperbolicity. The stable/unstable subspaces
\begin{subequations}\label{eq:decomposition-stable-unstable}
	\begin{empheq}[]{align}
	V^-_{\mathbf{x}(t)} =\;& \{ \mathbf{u} \in \mathbb{R}^N : \|\mathbf{Y}(\tau, \mathbf{x}(t))^{\phantom{-1}}\cdot\mathbf{u}\| \leq C\|\mathbf{u}\|e^{-\lambda \tau}, \forall \tau > t\},\\
	V^+_{\mathbf{x}(t)} =\;& \{ \mathbf{u} \in \mathbb{R}^N: \|\mathbf{Y}(\tau, \mathbf{x}(t))^{-1}\cdot\mathbf{u}\| \leq C\|\mathbf{u}\|e^{-\lambda \tau}, \forall \tau > t\},
	\end{empheq}
\end{subequations}
for some finite constant $C\geq 1$ and for some $\lambda > 0$, contain vectors in tangent space that decay/grow exponentially under the action of the linearised dynamics, while the neutral subspace
\begin{equation}
V^0_{\mathbf{x}(t)} = \{ \mathbf{u} \in \mathbb{R}^N : \mathbf{u} = a\mathbf{f}(\mathbf{x}(t)), \forall a \in\mathbb{R}\}
\end{equation}
contains vectors parallel to the local vector field. With these definitions, the three error components obey
\begin{equation}
\|\mathbf{e}^-(t)\| \leq C\|\mathbf{e}^-(0)\|e^{ -\lambda t}, \quad \|\mathbf{e}^+(t)\| \leq C\|\mathbf{e}^+(T)\|e^{\lambda (t - T)}\quad\mathrm{and}\quad\|\mathbf{e}^0(t)\| = |a_0(t)|\|\mathbf{f}(t)\|,
\end{equation}
suggesting that the shadowing error on the stable/unstable subspace decays/grows exponentially fast and is thus relevant only in the initial/final part of the time interval, while the error on the neutral subspace is distributed across the entire time span. By hyperbolicity assumption, the three subspaces are always transversal to each other. Hence, at any point in time, the magnitude of the three components can be bounded with the norm of $\|\mathbf{e}(t)\|$ as
\begin{equation}\label{eq:error-split-bounds}
	\|\mathbf{e}^-(t)\| \leq \frac{\|\mathbf{e}(t)\|}{\beta}, \quad \|\mathbf{e}^+(t)\| \leq \frac{\|\mathbf{e}(t)\|}{\beta}, \quad \mathrm{and} \quad |a_0(t)| \leq \frac{\|\mathbf{e}(t)\|}{\beta\|\mathbf{f}(t)\|},
\end{equation}
for some positive constant $\beta > 0$. Thus, the magnitude of the sensitivity error on the stable subspace can be written as
\begin{equation}\label{eq:error-decay-stable}
|\mathcal{E}^{T}_{1, -}(\mathbf{x}_0, p)| = \frac{1}{T} \int_0^T |J_{\partial \mathbf{x}}(t)^\top\cdot\mathbf{e}^-(t)|\,\mathrm{d}t
\leq \frac{1}{T}\int_0^T \|J_{\partial \mathbf{x}}(t)\|\|\mathbf{e}^-(t)\|\,\mathrm{d}t
\leq \frac{GC \|\mathbf{e}(0)\|}{ \beta T \lambda}[1 - e^{-\lambda T}],
\end{equation}
for some finite $G = \sup_t \|J_{\partial \mathbf{x}}(t)\|$. A similar calculation can be formulated for the sensitivity error associated to the unstable subspace, but now using the boundary conditions (\ref{eq:error-equations-b}) to obtain the bound \mbox{$\|\mathbf{e}(T)\| \leq \|\mathbf{e}(0)\| + 2B$}, to express $\|\mathbf{e}(T)\|$ as a function of $\|\mathbf{e}(0)\|$.

To obtain an expression for the sensitivity error associated to the neutral subspace, we first observe that direct substitution of $\mathbf{e}^0(t) = a(t)\mathbf{f}(t)$ into the differential equation (\ref{eq:error-equations-a}) leads to
\begin{equation}
a(t) = a_0 +\int_0^t T^0_{\mathrm{d}p}/T - \eta(s)\,\mathrm{d}s,
\end{equation}
with $a_0=a(0)$ a finite constant. Including the contribution from the time transformation, we find that
\begin{align}
	\mathcal{E}^{T}_{1,0}(\mathbf{x}_0, p) =&\; \frac{1}{T} \int_0^T J_{\partial \mathbf{x}}(t)^\top\cdot\mathbf{e}^0(t)\,\mathrm{d}t - \eta(t)[J(t) - \mathcal{J}^T] \,\mathrm{d}t.\nonumber \\
	=&\; \frac{a_0}{T} \int_{0}^T  \dot{J}(t)\,\mathrm{d}t +  \frac{1}{T}\int_0^T \dot{J}(t)\left[\int_0^t T^0_{\mathrm{d}p}/T - \eta(s)\;\mathrm{d}s\right] - \eta(t)[J(t) - \mathcal{J}^T]\;\mathrm{d}t, \label{eq:error-decay-neutral}
\end{align}
where we use the fact that $J_{\partial \mathbf{x}}^\top(t)\cdot\mathbf{f}(t) = \dot{J}(t)$. Using the decomposition (\ref{eq:transformation-decomposition}), we obtain
\begin{align}\label{eq:error-neutral}
\mathcal{E}^{T}_{1,0}(\mathbf{x}_0, p) =&  \frac{a_0}{T} \int_{0}^T  \dot{J}(t)\,\mathrm{d}t +  \frac{T^0_{\mathrm{d}p}/T - \bar{\eta}^\infty}{T}\int_0^T t \dot{J}(t) \mathrm{d}t \nonumber \\
& - \frac{1}{T}\int_0^T \dot{J}(t) \left[\int_0^t\tilde{\eta}^\infty(s)\mathrm{ds}\right] \mathrm{d}t - \frac{1}{T}\int_0^T \tilde{\eta}^\infty(t)\big[J(t) - \mathcal{J}^T\big]\mathrm{d}t \nonumber \\
=&  \frac{a_0}{T} \left[J(T) -J(0)\right] +  \left(\frac{T^0_{\mathrm{d}p}}{T} - \frac{1}{T}\int_0^T\eta(t)\mathrm{d}t\right)  \left[J(T) -  \mathcal{J}^T\right],
\end{align}
where integration by parts is used to drop the integral of the third term in the second step. Equation (\ref{eq:error-neutral}) shows that the sensitivity error along the stable subspace is made of two components. The first clearly decays as $1/T$, while the second contribution depends on the gradient ${T^0_{\mathrm{d}p}}/{T}$ and decays similarly if it can be shown that the difference in the parenthesis is
\begin{equation}\label{eq:C2-def}
\frac{T^0_{\mathrm{d}p}}{T} - \frac{1}{T}\int_0^T\eta(t)\mathrm{d}t = \frac{C_2^T(\mathbf{x}_0, p)}{T},
\end{equation}
for some constant $C_2^T(\mathbf{x}_0, p)$ that does not grow on average with $T$. In this case, using (\ref{eq:error-split-bounds}), the sensitivity error associated to the stable direction
\begin{align}
|\mathcal{E}^{T}_{1,0}(\mathbf{x}_0, p)| \leq& \frac{S}{T} \left(\frac{2\|\mathbf{e}(0)\|}{\beta \|\mathbf{f}(0)\|} + |C_2^T(\mathbf{x}_0, p)|\right),
\end{align} 
also asymptotically decreases as $1/T$, for some finite $S = \sup_t |J(t)|$. 

\subsection{Analysis of the terms $\|\mathbf{e}(0)\|$ and $C_2^T(\mathbf{x}_0, p)$}\label{sec:showing-e0}
Using the general solution (\ref{eq:general-solution-error-equation}), the error BVP (\ref{eq:error-equations}) can be transformed into the matrix equation
\begin{equation}\label{eq:error-bordered-system}
	\left[\begin{array}{c|c}
	\mathbf{M}(T, \mathbf{x}_0) & \mathbf{f}(T) \\[4pt]
	\hline\!\rule{0in}{.4cm} 
	\mathbf{f}^\top(0)               &  0\\
	\end{array} \right]\cdot
	\left[\begin{array}{c}
	\mathbf{e}(0) \\[4pt]
	\hline\!\rule{0in}{.4cm} C_{2}^T(\mathbf{x}_0, p)
	\end{array}\right] = 
	-\left[\begin{array}{c}
	\mathbf{r} \\[4pt]
	\hline\!\rule{0in}{.4cm}0
	\end{array}\right], 
\end{equation}
where we have defined for convenience the matrix $\mathbf{M}(T, \mathbf{x}_0) = \mathbf{Y}(T, \mathbf{x}_0) - \mathbf{I}$. The matrix on the left hand side, denoted in what follows as $\mathbf{B}(T, \mathbf{x}_0)$, has a bordered structure, which arises frequently in bifurcation analysis and continuation problems for dynamical systems \citep{Govaerts:2000vv}. Denoting by $\sigma_0(T, \mathbf{x}_0)$ the least singular value of this matrix, the bound
\begin{equation}
	\|\mathbf{e}(0)\|^2 + |C_2^T(\mathbf{x}_0, p)|^2  =  \|\mathbf{B}^{-1}(T, \mathbf{x}_0)\cdot\mathbf{r}\|^2 \leq \|\mathbf{B}^{-1}(T, \mathbf{x}_0)\|^2\;\|\mathbf{r}\|^2\leq\left(\frac{\|\mathbf{r}\|}{\sigma_0(T, \mathbf{x}_0)}\right)^2,
\end{equation}
can be obtained, leading to
\begin{equation}\label{eq:bounds-on-e-c2}
	\|\mathbf{e}(0)\| \leq \frac{2B}{\sigma_0(T, \mathbf{x}_0)},\quad  |C_2^T(\mathbf{x}_0, p)| \leq \frac{2B}{\sigma_0(T, \mathbf{x}_0)},
\end{equation}
where the definition (\ref{eq:bound-B}) has been used. Note that the bound on $C_2^T(\mathbf{x}_0, p)$ shows that the term of the left hand side of (\ref{eq:C2-def}) decays to zero as $T\rightarrow 0$. Summing now the three sensitivity error components, and using the above results, we finally obtain
\begin{align}\label{eq:sensitivity-error-1-final}
|\mathcal{E}^{T}_{1}(\mathbf{x}_0, p)| \leq& |\mathcal{E}^{T}_{1, -}(\mathbf{x}_0, p)| + |\mathcal{E}^{T}_{1, 0}(\mathbf{x}_0, p)| + |\mathcal{E}^{T}_{1, +}(\mathbf{x}_0, p)| \nonumber \\
\leq& \frac{1}{T}\bigg[\frac{4BS}{\sigma_0(T, \mathbf{x}_0)\beta \|\mathbf{f}(0)\|} +  \frac{2BS}{\sigma_0(T, \mathbf{x}_0)}+2GCB \frac{2/\sigma_0(T, \mathbf{x}_0)+1}{ \beta \lambda}[1 - e^{-\lambda T}] \bigg ].
\end{align} 

In equation (\ref{eq:sensitivity-error-1-final}) the norm of $\|\mathbf{f}(0)\|$ is generally positive on a chaotic trajectory, but can become small when the attractor includes an equilibrium point, like in the Lorenz equations. More importantly, the least singular value $\sigma_0(T, \mathbf{x}_0)$ can be arbitrarily small. In fact, we show in \ref{app:B-singular} that, for a given initial condition $\mathbf{x}_0$, the bordered system (\ref{eq:error-bordered-system}) is singular on a zero measure set of time spans $T^o_i$, $i=1, \ldots$, where
\begin{equation}\label{eq:chi-definition}
\mathbf{f}(0)^\top\cdot\mathbf{M}^{-1}(T^o_i, \mathbf{x}_0)\cdot\mathbf{f}(T^o_i) = 0, \quad i=1, \ldots.
\end{equation}
Around $T=T^o_i$, the least singular value of $\mathbf{B}(T, \mathbf{x}_0)$ behaves as $k_i|T-T^o_i|$ for some positive constant $k_i$, see \ref{app:svd-near-a-zero}. From this fact, we derive in \ref{app:power-law} that the probability density function of $1/\sigma_0(T, \mathbf{x}_0)$ features a power-law tail of the form $p(x) \simeq 1/x^n$, with $n=2$, for $1/\sigma_0(T, \mathbf{x}_0) \gg 1$. 

From a practical perspective, this implies that the probability density function of the sensitivity error $\mathcal{E}^{T}_{1}(\mathbf{x}_0, p)$ and of the sensitivity $\mathcal{J}^{T, \mathrm{P}}(\mathbf{x}_0, p)$ will display power-law tails with same exponent $n=2$. Heavy-tailed distributions have been observed in previous work related to sensitivity analysis of chaotic systems. For instance, \citep{Eyink:2004gk} reported power-law distributions of the adjoint gradients of finite-time averages obtained from the classical backward-in-time integration. In \citep{Wang:2013cx} it is suggested that shadowing-based sensitivity calculations might display heavy-tailed distributions. However, to the best of our knowledge, no statistical description of the sensitivity error for a shadowing-based algorithm has been previously reported.

Finally, we make a small note on the convergence of $T^0_{\mathrm{d}p}/T$. A useful consequence of the bound (\ref{eq:bounds-on-e-c2}) on $C_2^T(\mathbf{x}_0, p)$ is that dividing it by $T$, using the decomposition (\ref{eq:transformation-decomposition}) and rearranging, it can be obtained that
\begin{equation}\label{eq:Tdp-converges}
\frac{T^0_{\mathrm{d}p}}{T} = \bar{\eta}^\infty + \displaystyle\frac{C_2^T(\mathbf{x}_0, p)}{T} + \frac{1}{T}\int_{0}^{T} \tilde{\eta}^\infty(t)\,\mathrm{d}t,
\end{equation}
which shows that, as $T\rightarrow\infty$, ${T^0_{\mathrm{d}p}}/{T}$ converges correctly to $\bar{\eta}^\infty$. The convergence rate is initially $1/T$, because of the term $C_2^T(\mathbf{x}_0, p)/T $, but then transitions to $1/\sqrt{T}$, driven by the slower convergence of the finite-time average of $\tilde{\eta}^\infty(t)$. Numerical evidence for such convergence rates will be shown in section \ref{sec:numerical-example}.

\subsection{Discussion on the probability distribution of the sensitivity}\label{sec:discussconvergence}
For power law distributions of the form $p(x) \simeq 1/x^n$, central moments of order $m$ are undefined for $m\geq n - 1$. In the present case, with $n=2$, this implies that the mean and the variance of the sensitivity error will not converge as the periodic shadowing method is applied to an increasing number of trajectory segments, from independent initial conditions. Nevertheless, the overall convergence of the algorithm with $T\rightarrow\infty$ can still be shown from a practical perspective by replacing the mean and standard deviation of the sensitivity error with the median and interquartile range, respectively. These quantities are well defined for distributions with power-law tails and thus converge to finite values when the sensitivity algorithm is applied to an increasing number $M$ of independent trajectory segments, at a $1/\sqrt{M}$ rate. As a result, the median of the sensitivity error is proportional to the median of $1/\sigma_0$ divided by $T$. Since, the median of $1/\sigma_0$ is a bounded quantity, the median of the sensitivity error decays to zero as $T\rightarrow\infty$.

An alternative perspective is to consider the probability that the sensitivity error on a single trajectory segment of length $T$ is larger than a user-defined tolerance $\epsilon$. It can be shown that, for large $T\epsilon$, this probability is 
\begin{equation}
P(|\mathcal{E}^{T}_{1}| > \epsilon) \sim P\left({1}/{\sigma_0} > T \epsilon\right) \sim \int_{T \epsilon}^{\infty} {\sigma_0^{-2}} \ {\rm d}(1/\sigma_0)  \sim ({T \epsilon})^{-1}
\end{equation}
Thus, for a given $\epsilon$, $\lim_{T\rightarrow \infty} P(|\mathcal{E}^{T}_{1}| > \epsilon) = 0$.

\section{Numerical examples}\label{sec:numerical-example}
In this section, we demonstrate the method on the Lorenz equations \citep{Lorenz:1963tf}. Our objective is primarily to provide numerical evidence for the theoretical considerations of section \ref{app:error-analysis}. 

The Lorenz equations are
\begin{equation}\label{eq:lorenz-equations}
  \dot{\mathbf{x}}(t) = 
  	\begin{bmatrix}
	\dot{x_1}(t)\\
	\dot{x_2}(t)\\
	\dot{x_3}(t)\\
  \end{bmatrix} = 
  \begin{bmatrix}
   \sigma(x_2(t)-x_1(t))\\
   \rho x_1(t) - y_2(t) - x_1(t)x_3(t)/\gamma\\
   \gamma x_1(t)x_2(t)  - \beta x_3(t)
  \end{bmatrix} =
  \mathbf{f}(\mathbf{x}(t), \mathbf{p}),
\end{equation}
where $\mathbf{p} = (\rho, \sigma, \beta, \gamma)^\top$, and the standard parameters $\sigma=10$, $\beta=8/3$, $\rho = 28$ and $\gamma=1$ are used throughout, unless otherwise stated. The equations are parametrised by the additional parameter, $\gamma$, whose effect is discussed below. As in other studies on the Lorenz equations \citep{LEA:2000cr, Reick:2002co, Eyink:2004gk,Wang:2013cx,Liao:2016hn,Lasagna:2017tz}, we will consider the sensitivity of averages of the observable $J(\mathbf{x}(t), \mathbf{p}) = x_3(t)$ with respect to perturbations of the parameters $\rho$, $\beta$, and additionally, $\gamma$.

The linearised equation for (\ref{eq:lorenz-equations}) reads as
\begin{equation}\label{eq:lorenz-variational-equations}
  \dot{\mathbf{y}}(t) = 
  	\begin{bmatrix}
	\dot{y_1}(t)\\
	\dot{y_2}(t)\\
	\dot{y_3}(t)\\
  \end{bmatrix} = 
  \begin{bmatrix}
   -\sigma & \sigma & 0\\
   \rho - x_3(t)/\gamma  &    -1  & - x_1(t)/\gamma \\
  \gamma x_2(t)  & \gamma x_1(t)  & -\beta\\
  \end{bmatrix}\cdot
  \begin{bmatrix}
	y_1(t)\\
	y_2(t)\\
	y_3(t)\\
  \end{bmatrix} =
  \mathbf{f}_{\partial \mathbf{x}}(\mathbf{x}(t), \mathbf{p})\cdot\mathbf{y}(t).
\end{equation}
The forcing functions for the non-homogenous sensitivity equations (\ref{eq:forward-sensitivity-problem-a}) are 
\begin{align}
	&\mathbf{f}_{\partial \rho}(t) = [0, x_1(t), 0]^\top,\; \mathbf{f}_{\partial \beta}(t) = [0, 0, -x_3(t)]^\top\;\mathrm{and}\;\, \mathbf{f}_{\partial \gamma}(t) = [0, x_1(t)x_3(t)/\gamma^2, x_1(t)x_2(t)]^\top,
\end{align}
respectively, while $J_{\partial \mathbf{p}}(t) = [0, 0, 0]^\top$ and $J_{\partial \mathbf{x}}(t) = [0, 0, 1]^\top$.

\begin{figure}[tbp]
	\centering
	\includegraphics[width=0.9\textwidth]{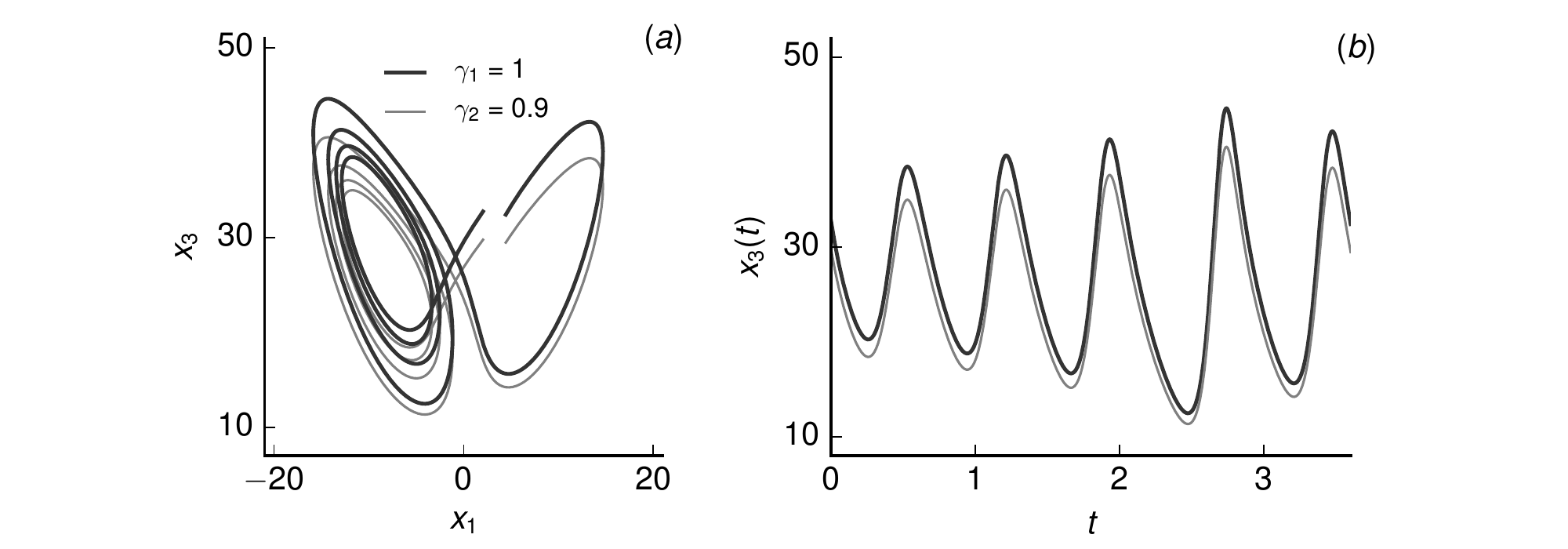}
	\caption{Effect of the parameter $\gamma$ on a short trajectory of the Lorenz equations, starting from a point on the attractor. The initial condition of the trajectory for $\gamma_2 = 0.9$ is obtained by applying the transformation (\ref{eq:stretch-map}) to the initial condition of the trajectory for $\gamma_1=1$. }
	\label{fig:stretch-attractor-deformation}
\end{figure}

The equations (\ref{eq:lorenz-equations}) are different from classical definitions in that they are parametrised by an additional parameter, $\gamma$. This parameter describes the state evolution under the coordinate transformation $\mathbf{h}_\gamma$ defined by
\begin{equation}\label{eq:stretch-map}
	\mathbf{h}_\gamma : ({x}_1, {x}_2, {x}_3) \rightarrow (x_1, x_2, \gamma x_3).
\end{equation}
As illustrated in figure \ref{fig:stretch-attractor-deformation}, for $\gamma > 1$ the phase space is stretched in the vertical direction, while it is compressed if $\gamma < 1$. For $\gamma=1$, the reference value, the standard equations are obtained. Varying $\gamma$ produces an up/down-ward stretch of the attractor and a direct change in the statistics involving the coordinate $x_3(t)$. 

The reason why we consider such case is that structural perturbations of the equations that are equivalent to smooth coordinate transformations do not drive the system into bifurcations: equilibria, periodic orbits as well as more complicated attractors remain topologically unchanged as $\gamma$ is varied. This situation parallels the case of hyperbolic systems where, for small perturbations of the parameters, the perturbed system is topologically conjugate to the original system (see pg. 38 of Ref.~\citep{Guckenheimer:up}). Such type of perturbation is also considered in Ref.~\citep{Chater:2017dw} for the Lorenz equation and, earlier, for maps in Ref.~\citep{Grossmann:1984kh}.

Most importantly, the motivation to consider this problem is that the shadowing direction is known explicitly for this case, enabling a detailed verification of the theoretical predictions. Direct substitution in (\ref{eq:lorenz-variational-equations}) shows that, for a trajectory $[x(t), y(t), z(t)]$ satisfying (\ref{eq:lorenz-equations}), the shadowing direction
\begin{equation}\label{eq:analytic-shadowing-direction}
\mathbf{y}_{\mathrm{S}}(t) = \left[0, 0, z(t)\right]^\top	
\end{equation}
is the solution of the sensitivity equations at $\gamma=1$ and with $\eta(t) = 0$ because the stretching does not affect the temporal dynamics of the problem. Note that considering (\ref{eq:analytic-shadowing-direction}) instead of the ``canonical'' shadowing direction satisfying the orthogonality condition (\ref{eq:orthogonality-constraint-shadow}) does not affect the predictions of the error analysis. The sensitivity of the average can thus be found analytically
\begin{equation}\label{eq:stretch-asymptotic-sensitivity}
	\mathcal{J}_{\partial \gamma}^T(\mathbf{x}_0, \gamma) = \frac{1}{T}\int_0^T J_{\partial \mathbf{x}}(t)^\top\cdot\mathbf{y}(t)\,\mathrm{d}t = \frac{1}{T}\int_0^T x_3(t)\,\mathrm{d}t = \mathcal{J}^T(\mathbf{x}_0, \gamma).
\end{equation}

We also study the classical problem where the sensitivity of the observable $x_3(t)$ with respect to the parameter $\rho$ is of interest. Although the qualitative effect of $\rho$ on the attractor is roughly similar to that of $\gamma$, there are important differences between the parameters. As it is known from the investigations of Sparrow (\citep{Sparrow:2012hi}), perturbations of $\rho$ induce homoclinic bifurcations, i.e.~some unstable periodic orbits passing very near the unstable equilibrium at the origin of (\ref{eq:lorenz-equations}) can appear/disappear upon small perturbations of $\rho$. Hence, the Lorenz equations with standard parameters are not strictly hyperbolic. This class of systems is sometimes referred to in the literature as ``quasi-hyperbolic'', or ``singularly-hyperbolic'', a weaker definition of hyperbolicity \citep{Viana:2000bi}. The effect of this feature on the sensitivity results and the difference with the analysis of the parameter $\gamma$ will be illustrated in detail in the next sections.

Numerical integration of the nonlinear and linearised equations is performed using a classical fourth-order accurate Runge-Kutta method. We use $dt=10^{-2}$ for the nonlinear simulations, as this is sufficient to achieve time step independence of the long-time statistics. For the linearised equations we used a shorter time step, $dt=10^{-3}$, to obtain accurate solutions of (\ref{eq:forward-sensitivity-problem}). The linearised equations are solved in a coupled manner with the nonlinear equations by propagating forward in time the augmented system. Numerical quadrature of all time integrals, e.g. in (\ref{eq:sensitivity-integral-adjoint}) or (\ref{eq:sensitivity-limit-1}), is performed by augmenting the equations with a quadrature equation so that integration maintains the same order of accuracy as the time stepping. All the numerical results reported in this section are obtained using the tangent algorithm. 

\subsection{Singularity conditions}\label{eq:singularity-conditions}
We first focus on the spectral properties of the matrices $\mathbf{B}(T, \mathbf{x}_0)$ and $\mathbf{M}(T, \mathbf{x}_0)$ to provide numerical evidence for some of the statements made in section \ref{app:error-analysis}. 
\begin{figure}[tbp]
	\centering
	\includegraphics[width=0.95\textwidth]{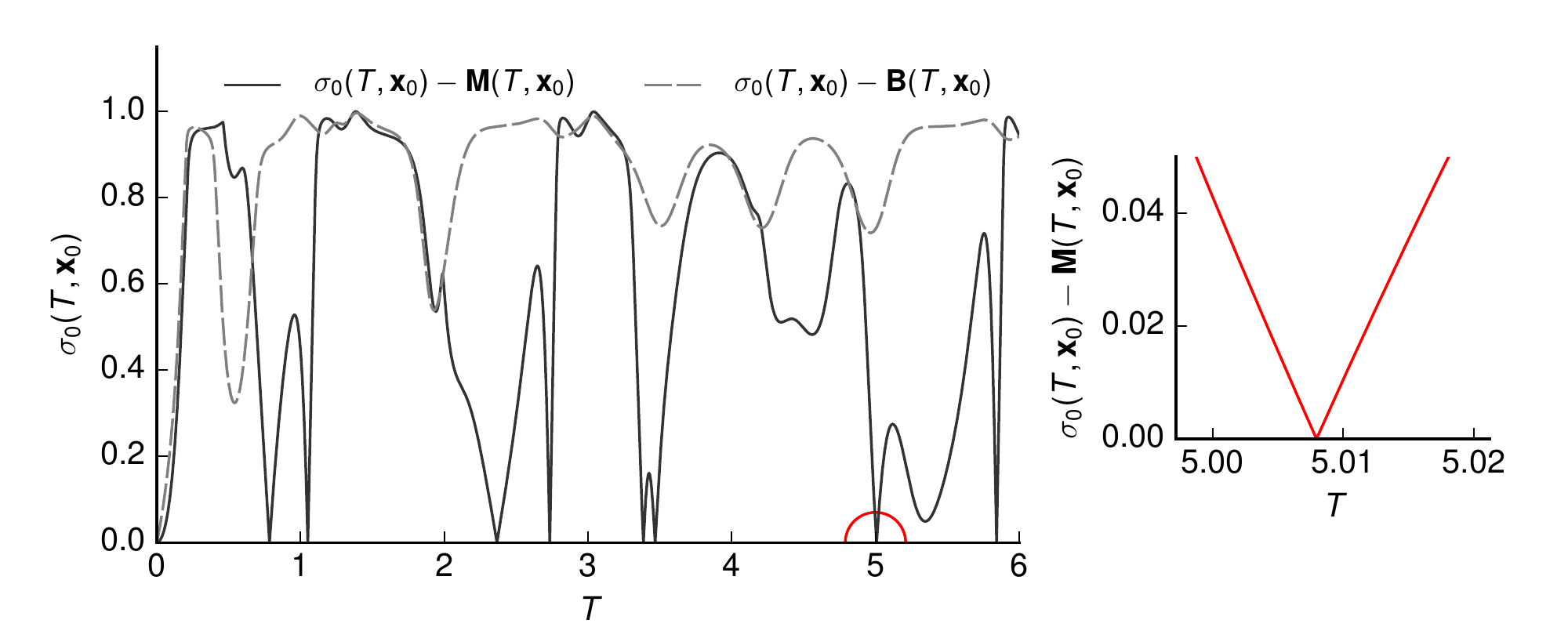}
	\caption{The least singular value of the matrices $\mathbf{M}(T, \mathbf{x}_0)$ (solid line) and  $\mathbf{B}(T, \mathbf{x}_0)$ (dashed line) as a function of the time span $T$, for some initial condition $\mathbf{x}_0$ lying on the attractor. The inset on the right hand side of the figure illustrates how the least singular value of $\mathbf{M}(T, \mathbf{x}_0)$ approaches zero near $T=5$.}
	\label{fig:singular-values-of-bordered-nonbordered-systems-just-a-time-history}
\end{figure}
Figure \ref{fig:singular-values-of-bordered-nonbordered-systems-just-a-time-history} shows the behaviour of the least singular value of these two matrices, where we take one initial condition $\mathbf{x}_0$ on the attractor and study the effect of the time span $T$.

For $T=0$, both matrices are singular since $\mathbf{Y}(0, \mathbf{x}_0) = \mathbf{I}$. As $T$ increases, the matrix $\mathbf{M}(T, \mathbf{x}_0)$ becomes repeatedly singular at certain time spans. Close to singularity points (around some $T^o$), the least singular value of $\mathbf{M}(T, \mathbf{x}_0)$ behaves as $k|T-T^o|$, for some positive constant $k$, as shown in the smaller panel on the right hand side, focusing around $T\sim 5$, and as predicted by the analysis of \ref{app:svd-near-a-zero}. On the other hand, the least singular value of the bordered matrix $\mathbf{B}(0, \mathbf{x}_0)$ never approaches zero but instead fluctuates around one and does not change, in a statistical sense, when $T$ increases, e.g. it does not get asymptotically smaller. 

\begin{figure}[tbp]
	\centering
	\includegraphics[width=0.98\textwidth]{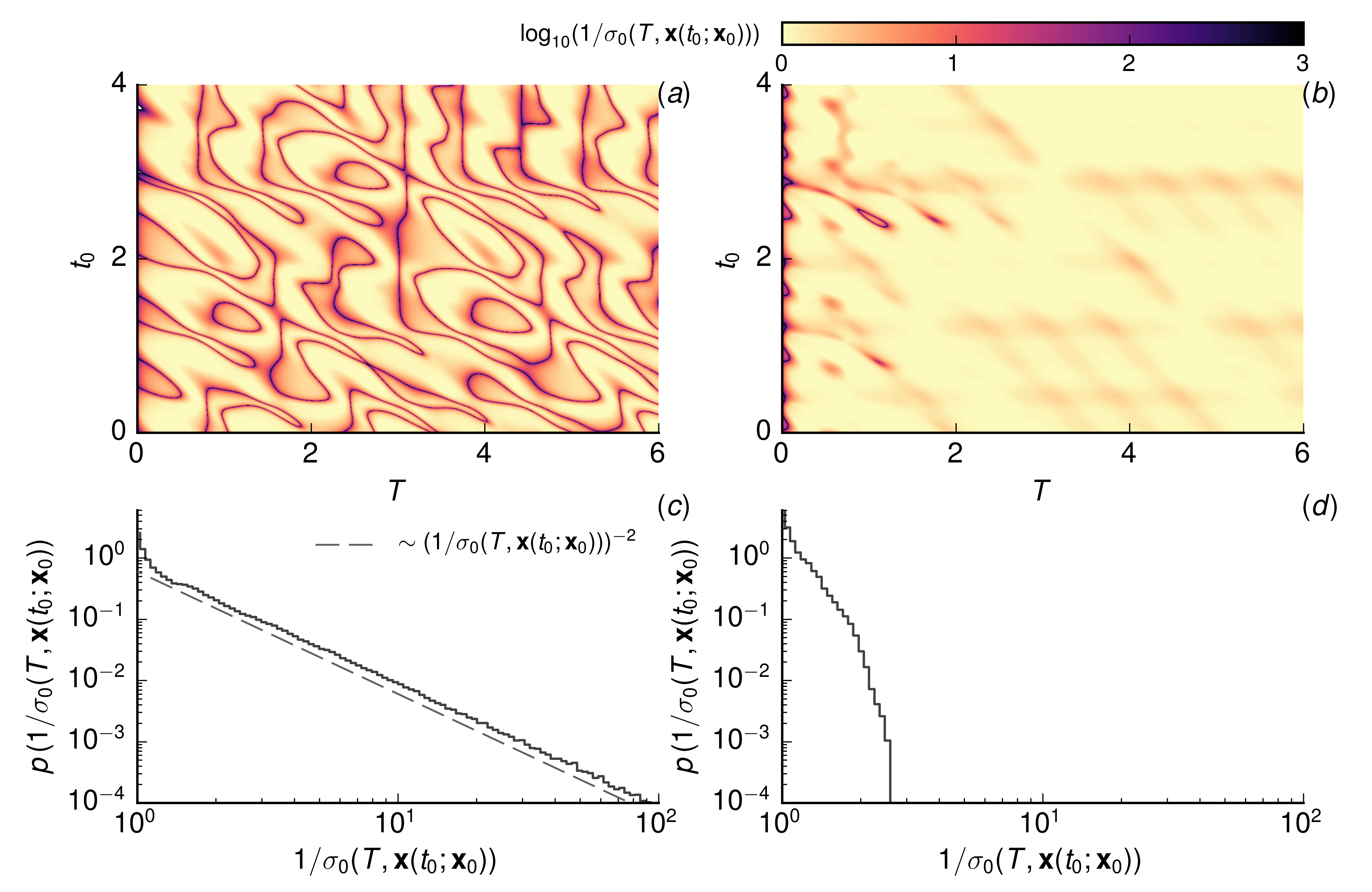}
	\caption{The inverse of the least singular value of the matrices $\mathbf{M}(T, \mathbf{x}_0)$ and $\mathbf{B}(T, \mathbf{x}_0)$, panels (a) and (b), respectively. Panels (c) and (d) show the probability density function of the inverse of the least singular value for the same two matrices as in panels (a) and (b).} 
	\label{fig:singular-values-of-bordered-nonbordered-systems}
\end{figure}

Changing the initial condition $\mathbf{x}_0$ does not change the essence of the results of figure \ref{fig:singular-values-of-bordered-nonbordered-systems-just-a-time-history}. This is studied by propagating the original initial condition $\mathbf{x}_0$ forward in time under the dynamics by a time $t_0$ and repeating the study of figure \ref{fig:singular-values-of-bordered-nonbordered-systems-just-a-time-history}. The results are reported in panels (a) and (b) of figure \ref{fig:singular-values-of-bordered-nonbordered-systems}. To better highlight the singularity, the colour map denotes the base ten logarithm of the inverse of $\sigma_0$. The figure focuses on a rather short range of $T$ and $t_0$, but the data is statistically homogeneous in $T$ and $t_0$ for $T$ larger than about 2 to 3 time units. One important observation from this analysis is that the least singular value of $\mathbf{B}(T, \mathbf{x}_0)$ does not seem to approach zero for $T$ larger than about two time units. 
Panels (c) and (d) of figure \ref{fig:singular-values-of-bordered-nonbordered-systems} report the probability density function of the inverse of the least singular value of the matrices $\mathbf{M}(T, \mathbf{x}_0)$ and $\mathbf{B}(T, \mathbf{x}_0)$, respectively. These distributions are constructed by sampling $\sigma_0(T, \mathbf{x}_0)$ for $T=5$ for initial conditions on the attractor. It can be observed that the probability density function of the inverse of the least singular value of $\mathbf{M}(T, \mathbf{x}_0)$ displays a power-law tail of the form $p(1/\sigma_0)\simeq 1/\sigma_0^{2}$, as predicted by the analysis reported in \ref{app:power-law}. On the other hand the distribution of the inverse of the least singular value of $\mathbf{B}(T, \mathbf{x}_0)$ does not display the same distribution, but falls off quite rapidly. This seems to be a particular feature of the Lorenz equations, and is not generally true for other systems, as shown by the example in \ref{sec:aero-elastic}, a chaotic aero-elastic system.

\subsection{Sensitivity with respect to perturbations to $\gamma$}\label{sec:gamma-sensitivity}
We first consider the statistics of the error on the sensitivity of the observable $J(\mathbf{x}(t), \mathbf{p}) = x_3(t)$ with respect to the parameter $\gamma$, at $\gamma=1$.
\begin{figure}[htbp]
	\centering
	\includegraphics[width=0.95\textwidth]{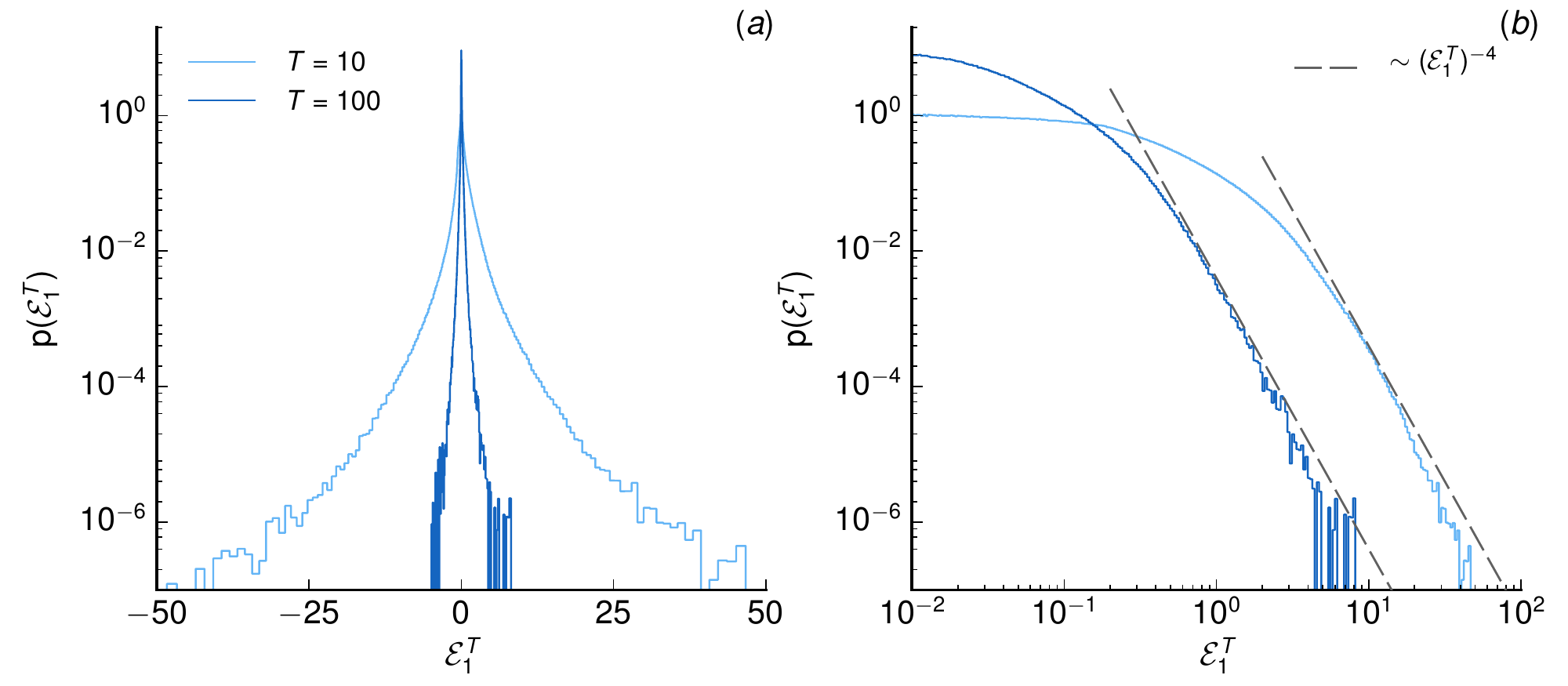}
	\caption{Distribution of the sensitivity error for increasing time spans. The right panel show a close-up of the right tail of the full distribution in the left panel.}
	\label{fig:stretch-error-histograms}
\end{figure}
Using periodic shadowing, we obtained about ten millions samples of the gradient $\mathcal{J}^{T, \mathrm{P}}_{\partial \gamma}(\mathbf{x}_0)$ from independent initial conditions $\mathbf{x}_0$ on the attractor, and calculated the error (\ref{eq:sensitivity-error-1}) directly for $T=10$ and $T=100$. Results are reported in figure \ref{fig:stretch-error-histograms}, showing the probability density function of the shadowing error (\ref{eq:sensitivity-error-1}). The left panel shows the full distribution, while the right panel focuses on the right tail, in a log-log plot to highlight the asymptotic trend. It can be observed that the probability density function displays tails falling at least as fast as $(\mathcal{E}_1^T)^{-4}$ and probably faster for large $\mathcal{E}_1^T$, much steeper than the predicted $(\mathcal{E}_1^T)^{-2}$. This is a consequence of the fact that the least singular value of the bordered matrix $\mathbf{B}(T, \mathbf{x}_0)$ does not approach zero, but appears to remain bounded. Analyses not reported here show that the distribution of the sensitivity $\mathcal{J}^{T, \mathrm{P}}_{\mathrm{d}\gamma}$ has similar tails. 

\begin{figure}[htbp] 
	\centering
	\includegraphics[width=1\textwidth]{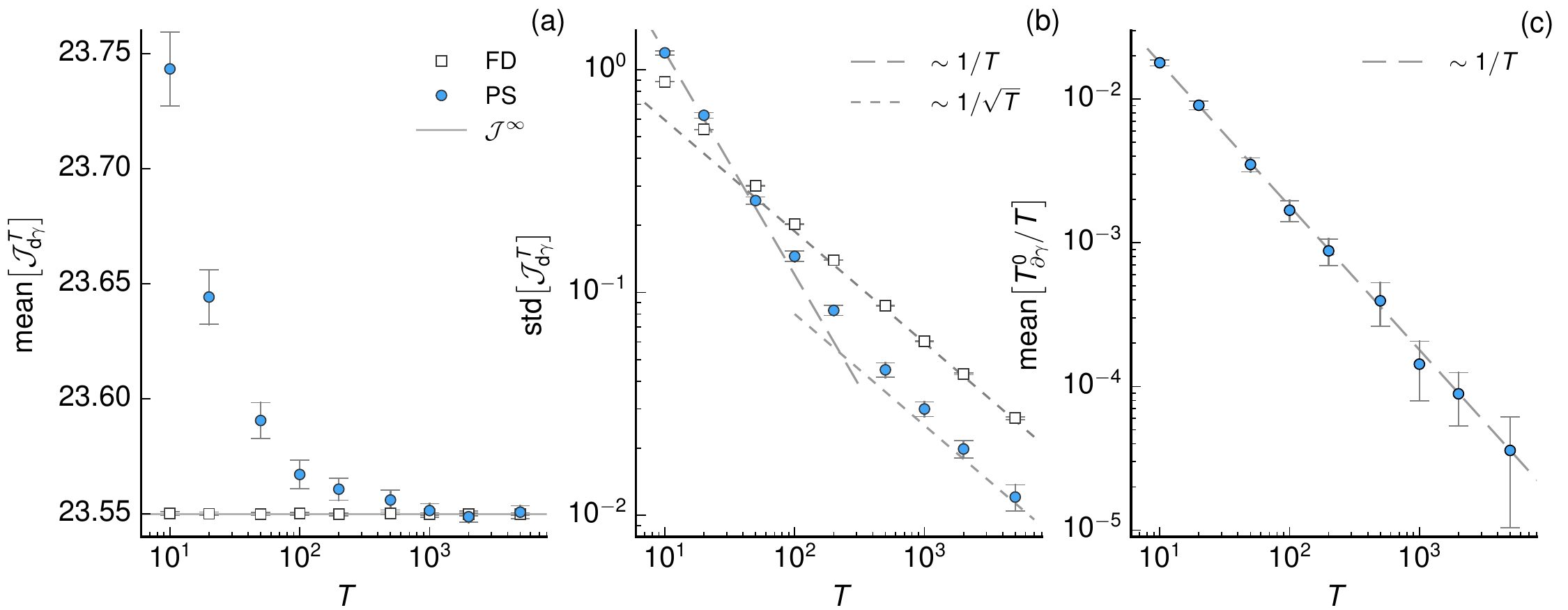}
	\caption{Mean, (a), and standard deviation, (b), of the sensitivity $\mathcal{J}_{\mathrm{d} \gamma}^T$ computed from multiple repetitions of the periodic shadowing (PS, blue circles) and finite-difference algorithms (FD, white squares) from independent initial conditions, as a function of the time span $T$. The dashed lines in panel (b) define the predicted error scaling. Panel (c) shows the average gradient $T_{\partial \gamma}/T$ as a function of $T$.}
	\label{fig:stretch-convergence}
\end{figure}

We then compare periodic shadowing sensitivity calculations with finite difference (FD) approximations of the gradient, as a function of the time span $T$. For each $T$, the periodic shadowing method is applied to consecutive trajectory segments lying on the attractor. The FD approximations are obtained using a centred second-order scheme, with $\Delta \gamma=0.5$, using averages over trajectory segments of length $T$ lying on the attractors at $\gamma=1\pm\Delta \gamma$. When equilibrium trajectory segments are used, the mean of the FD approximations does not depend on $T$, but the standard deviation does and decays asymptotically as $1/\sqrt{T}$. We repeat the algorithms for a number of times that is sufficient to bring the standard error bars down to a level that enables trends to be extracted accurately (hundreds of thousands repetitions is not uncommon for short spans with larger variance -- error bars define plus/minus three times the standard error on the mean or standard deviation \citep{Rao:2009tq}). 
Because of the fast drop of the probability distributions of $\mathcal{J}_{\mathrm{d} \gamma}^T$, both the mean and standard deviation converge as the number of samples is increased. We therefore calculate and plot these two quantities, rather than the median and the interquartile range. This would have not be possible in the general case where the gradient is distributed according to a heavy-tailed distribution. Results are reported in figure \ref{fig:stretch-convergence}. Panel (a) shows the mean sensitivity, and panel (b) the sample standard deviation. The dashed lines in panel (b) denote the expected scaling of the standard deviation from the error analysis, up to constants that have been adjusted to match the data points. Panel (c) shows the mean gradient $T_{\mathrm{d}\gamma}/T$.

The results show that, as $T\rightarrow\infty$, the expectation of the periodic shadowing sensitivity converges to the correct value, the long-time average of the observable $\mathcal{J}^\infty$, as defined in (\ref{eq:stretch-asymptotic-sensitivity}), and denoted in panel (a) by the horizontal line (this is obtained from a chaotic simulation with $T=10^7$). This is the same value obtained from the finite difference gradient approximations. The sample standard deviation of the periodic shadowing sensitivity calculations follows the scaling predicted by the error analysis in section \ref{app:error-analysis}. For short time spans, $T\lesssim200$, the sample standard deviation decays as $1/T$, as the shadowing error contribution (\ref{eq:first-thing-to-show}) dominates the variability across initial conditions. For larger $T$, the standard deviation decays as $1/\sqrt{T}$ since the contribution (\ref{eq:random-error}) associated to the finite-time averaging dominates. This is also the asymptotic decay rate of the standard deviation of the FD gradients. Analysis of the standard deviation associated to the evaluation of the integrals such as (\ref{eq:sensitivity-error-1}) shows that the standard deviation (denoted as $\mathrm{std}[]$) of the centred, second-order accurate FD gradient is proportional to $\mathrm{std}[x_3]/\Delta\gamma/\sqrt{T}$, while that of the periodic shadowing calculations, for large $T$, is $\mathrm{std}[x_3]/\sqrt{T}$, half of that of the FD gradients in the present case where $\Delta\gamma=0.5$. This is indeed observed in panel (b).

To conclude this section, we observe that, as indicated in section \ref{sec:showing-e0}, the ratio $T_{\mathrm{d}\gamma}^0/T$ must decay to $\bar{\eta}^\infty$ when $T\rightarrow\infty$. In the present case, the state space stretching induced by the parameter $\gamma$ does not change the time scale of the system, and $\bar{\eta}^\infty =0$. This is indeed observed in our calculations, as seen in figure \ref{fig:stretch-convergence}-(c).

\subsection{Sensitivity with respect to perturbations to $\rho$}
We now discuss sensitivity results of averages of the same observable with respect to the parameter $\rho$, for increasing time spans $T$. We compare periodic shadowing sensitivities with a) sensitivity calculations on unstable periodic orbits (UPOs) following Ref.~\citep{Lasagna:2017tz}, b) finite-difference gradients (with $\Delta\rho=0.5$), and c) Least-Squares Shadowing gradient calculations. The LSS data is obtained from digitizing data points reported in figure 6 of Ref.~\citep{Wang:2014hu} for the Lorenz equations at the same parameter values. Note that the LSS statistics are obtained from 10 repetitions of the algorithm. To increase confidence in our results and extract precise trends, statistics for periodic shadowing, UPOs and finite-difference calculations, are obtained by repeating each algorithm for a number of times sufficient to bring the standard error bars down to a size comparable to that of the symbols in the graph (e.g. for $T=100$ we obtained $1.1\times10^3$, $1.7\times10^4$ and $1.1\times10^5$ independent samples for UPOs, periodic shadowing and FD, respectively). The sensitivity calculations on UPOs were performed using the method discussed in Ref.~\citep{Lasagna:2017tz}. Since periodic orbits have a fixed period, statistics are calculated using orbits with period variation not greater than one time unit. 

\begin{figure}[htbp]
	\centering
	\includegraphics[width=0.95\textwidth]{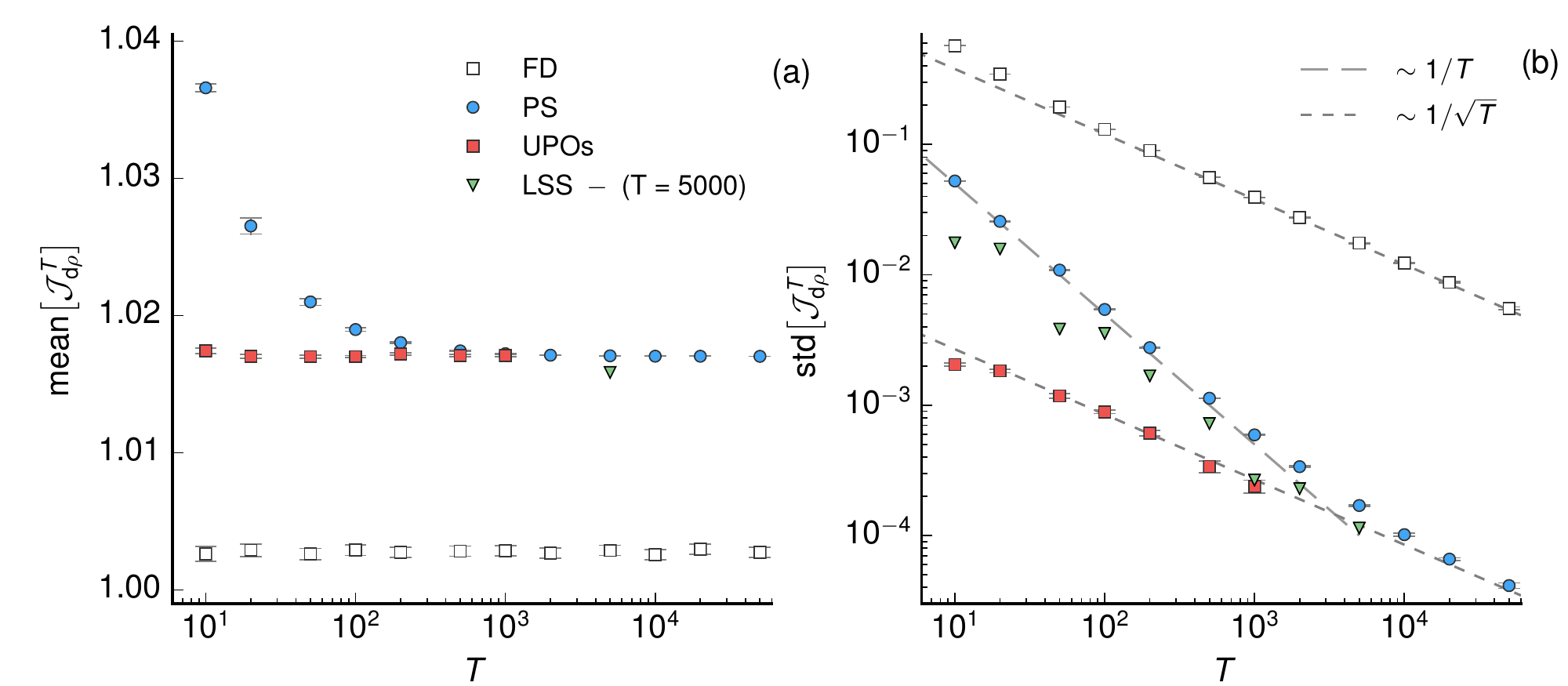}
	\caption{Arithmetic average, (a), and standard deviation, (b), of the sensitivity $\mathcal{J}^T_{\mathrm{d} \rho}$ computed from multiple repetitions of various algorithms, i.e.~ periodic shadowing (PS), sensitivity calculations on unstable periodic orbits (UPOs), finite-difference approximations (FD) and Least Squares Shadowing (LSS). For LSS data points have been digitised from figure 6 in \citep{Wang:2014hu}.}
	\label{fig:convergence}
\end{figure}

Results are reported in figure \ref{fig:convergence}. The arithmetic average of the sensitivity obtained from multiple repetitions of the various algorithms is reported in panel (a), while panel (b) shows the sample standard deviation. It can be observed that the arithmetic average of the periodic shadowing sensitivity converges to a value around $\mathcal{J}^T_{\mathrm{d} \rho} \simeq 1.017$ as $T$ is increased. Different variants of LSS produce similar values gradients \citep{Wang:2014hu,Liao:2016hn}. The sensitivity calculated from UPOs, which is not affected by the shadowing error $\mathcal{E}_1^T$ also converges, on average, to such a value. The data point for LSS at $T=5000$ lies also close to this value \citep{Wang:2014hu}. The important feature of figure \ref{fig:convergence}-(a) is that the finite-difference approximation of the gradient is significantly lower than what predicted by all the other methods. This is not a random error, but a reproducible bias in the average value over hundreds of thousand repetitions of the various algorithms from different initial conditions on the attractor. We have carefully checked that this bias is independent from the step size for numerical integration of the nonlinear equations or the step $\Delta\rho$ used for the finite difference approximation. The same bias has been already observed and discussed in Ref.~\citep{Lasagna:2017tz}. Discussion follows below.

\begin{figure}[tbp]
	\centering
	\includegraphics[width=0.95\textwidth]{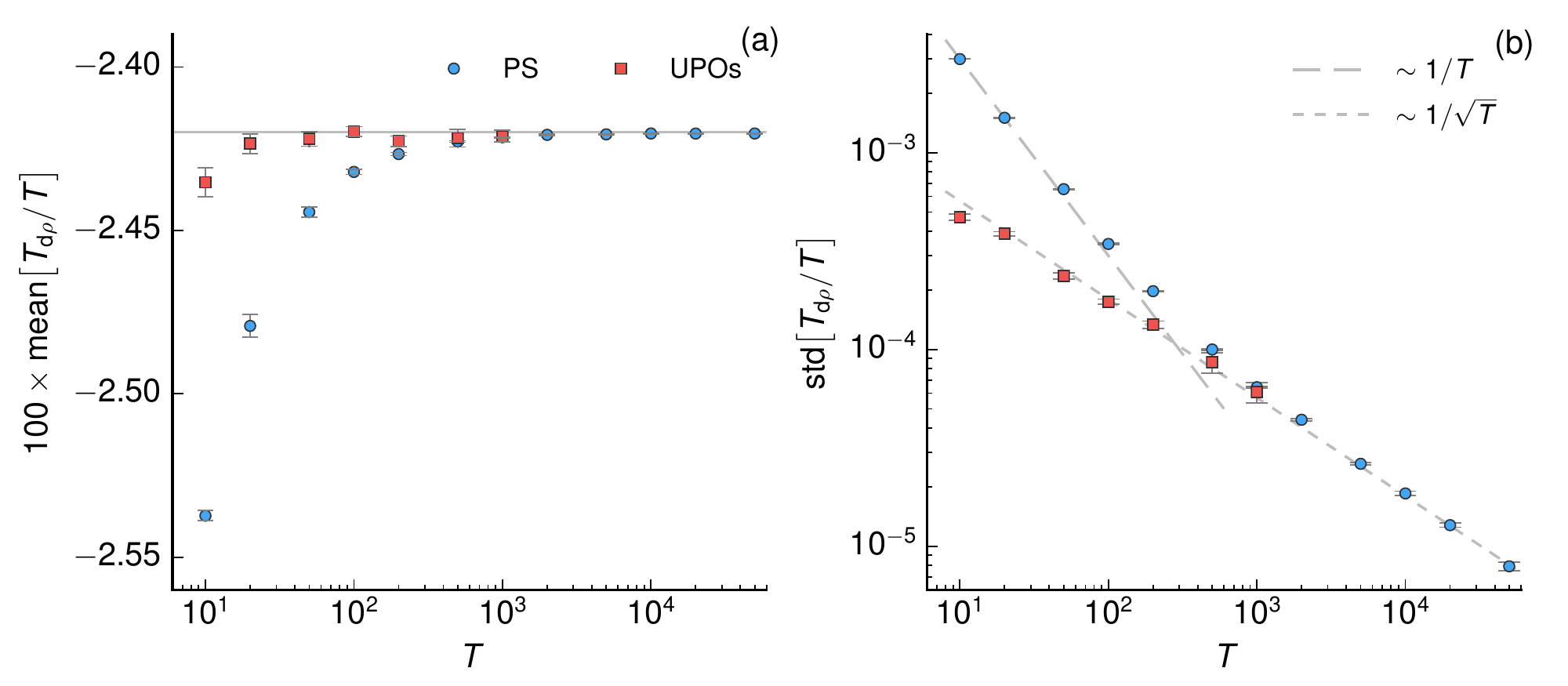}
	\caption{Arithmetic average, (a), and standard deviation, (b), of the period gradient computed from multiple periodic shadowing (PS) sensitivity calculations as well as from unstable periodic orbits (UPOs).}
	\label{fig:convergence-with-T-TpT}
\end{figure}

As predicted, the sample standard deviation of the sensitivity calculated with the periodic shadowing algorithm, panel (b), initially decays as $1/T$ and asymptotically as $1/\sqrt{T}$, similarly to that of LSS. The threshold at which the decay rate changes to $1/\sqrt{T}$ is around 5000 time units. This seem to be the same for the LSS data points. It is quite remarkable that this threshold is orders of magnitude larger than the typical time scale of the dynamics (the shortest UPO has period $\approx 1.55$ time units), and higher than what observed in figure \ref{fig:stretch-convergence} for the sensitivity with respect to $\gamma$. For periods longer than this threshold, the standard deviation is comparable to that computed over the UPOs, as the error $\mathcal{E}_0^T$ dominates. Note that, as discussed in section \ref{app:error-analysis}, for UPOs the standard deviation already decays as $1/\sqrt{T}$ from short time spans. Analysis on the statistics of the gradient $\mathcal{J}^T_{\mathrm{d}\rho}$, not reported here for the sake of brevity, show that this quantity, as well as the gradient $T_{\mathrm{d}\rho}/T$ is distributed according to a distribution with tails similar to those observed in figure \ref{fig:stretch-error-histograms} for the sensitivity with respect to the parameter $\gamma$.

A further point of interest is that the standard deviation of the FD gradients is more than two orders of magnitude higher than that from UPOs. The same analysis reported for the previous case shows that the standard deviation of the centred finite-difference gradient is proportional to $\mathrm{std}[x_3]/\Delta\rho/\sqrt{T}$, while that of UPOs is proportional to $\mathrm{std}[(x_3)_{\mathrm{d}\rho}]/\sqrt{T}$. The difference between the two is thus not just due to $\Delta\rho$ (here $\Delta \rho = \pm 0.5$), but primarily to the standard deviation of the quantities under the average, where $(x_3)_{\mathrm{d}\rho}(t)$, solution of the BVP (\ref{eq:forward-sensitivity-problem}) is of order 1, while the $x_3(t)$ spans the full attractor and $\mathrm{std}[x_3] \approx 8.6$ at the standard parameters.

Figure \ref{fig:convergence-with-T-TpT} shows statistics of the gradient $T_{\mathrm{d}\rho}/T$ as a function of the time span $T$. In panel (a) the mean gradient is reported, while the sample standard deviation is reported in panel (b). The figure shows data for periodic shadowing calculations as well for UPOs. Note that for UPOs, the gradient $T_{\mathrm{d}\rho}/T$ cannot be set arbitrarily, but is the unique value that allows the perturbed orbit to remain periodic upon a parameter perturbation (see details in \citep{Lasagna:2017tz}). The key result of figure \ref{fig:convergence-with-T-TpT} is that the gradient $T_{\mathrm{d}\rho}/T$ converges, as $T$ is increased, to a well defined value, about $-2.42\times10^{-2}$, indicated in panel (a) by the horizontal line. This value is the same obtained from periodic orbits, where there is no shadowing error and where the gradient has a well defined meaning, as previously suggested. Analysis of the standard deviation of the periodic shadowing calculations in panel (b) shows that convergence to the asymptotic value is achieved at a $1/T$ rate initially and then at a $1/\sqrt{T}$ rate for longer time spans, as predicted in the error analysis section. 

\begin{figure}[tbp]
	\centering
	\includegraphics[width=0.95\textwidth]{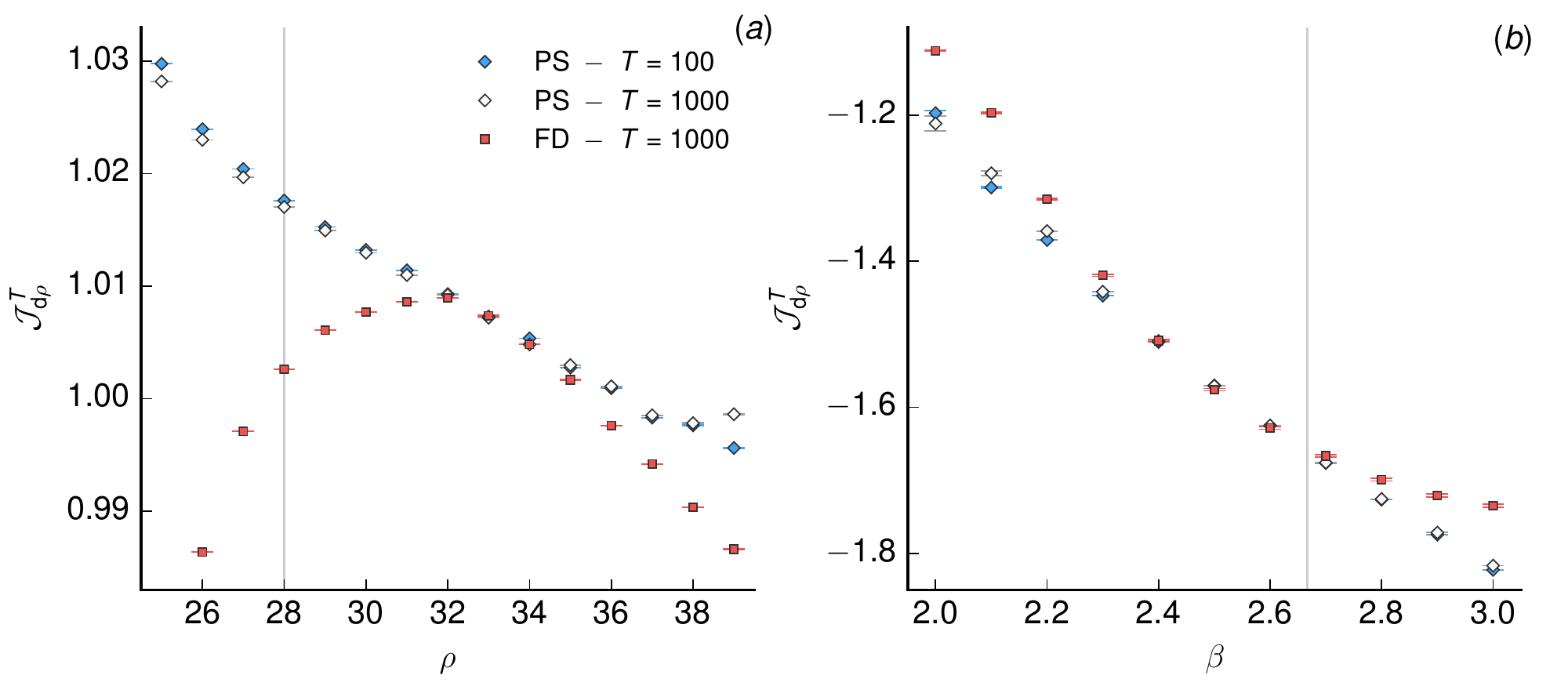}
	\caption{Sensitivity calculations of the observable $J(\mathbf{x})=x_3$ with respect to $\rho$, panel (a), and $\beta$, panel (b), as a function of the same two parameters, using periodic shadowing (PS) and finite-difference approximations $\mathrm{FD}$. The vertical line define the standard parameter values.}
	\label{fig:varying-parameters-comparison-FD}
\end{figure}

As shown in figure \ref{fig:convergence}-(a), there is a clear difference between the FD sensitivity approximation and all the other linear, shadowing-based sensitivity methods. To obtain a better insight into this bias, we performed sensitivity calculations for other parameter settings. We calculated the sensitivity of the same observable with respect to $\rho$ using periodic shadowing ($T=100$ and 1000) and FD ($T=1000$) for different values of $\rho$ in the interval $[25, 39]$, with $\beta=8/3$. Note that chaotic behaviour occurs for $\rho \geq 24.74$. We also calculated the sensitivity of the same observable with respect to the parameter $\beta$, for several values in the range $[2, 3]$, for $\rho=28$. For each parameter setting, we repeated the sensitivity calculations over a sufficient number of independent trajectory segments on the attractor to accurately extract the underlying trends.

The results, reported in figure \ref{fig:varying-parameters-comparison-FD}, indicate that the bias is a complex function of the parameters. For some parameter combinations, e.g. $\rho=33, \beta=8/3$, it is as small as the standard error bars, while for others, e.g. low $\rho$, the bias gets larger. The second observation is that the FD gradient approximation appears to be a smooth function of the parameters: not only is the average of the observable a smooth function of $\rho$ and $\beta$, but its derivative with respect to these parameters is a smooth function too.

\subsection{Discussion on the bias error}\label{sec:discussion}
As previously discussed, it is well known that the Lorenz system is not uniformly hyperbolic at the standard parameters \citep{Viana:2000bi}. For non-hyperbolic systems, the problem is the lack of structural stability \citep{Robert:2000fn}, so that the attracting set is always infinitesimally close to bifurcations \citep{Reick:2002co}. Upon a small structural perturbation the structure of the attractor can collapse or change suddenly, with macroscopic changes in the qualitative properties of the motion, thus rendering statistical quantities such as (\ref{eq:long-time-average}) discontinuous, and hence not differentiable with the parameters. This is well known for low dimensional systems, \citep{Ershov:1993el, Farmer:1985iv, Robert:2000fn}, but expected to be a general feature of many physical systems, where hyperbolicity is the exception and not the norm. In such situations, the limit (\ref{eq:derivative-as-limit}) does not itself formally exist, as the infinite time average is not a continuous function of the parameters. Note that empirical observations on the Lorenz equations, such as those in reported in this paper, suggest that statistics appear as if they were smooth functions of $\rho$. As suggested in Ref.~\citep{Reick:2002co}, investigating the predictions of linear response theory on the Lorenz equations, some observables might still behave continuously across such bifurcations.

In these conditions, sensitivity analysis of statistical quantities using linear shadowing-based methods such as periodic shadowing, LSS or using UPOs can be questionable. The linear problem (\ref{eq:forward-sensitivity-problem}) can be solved regardless of the hyperbolicity characteristics of the system at hand. Our interpretation, however, is that structural changes of the attractor under finite parameter perturbations might imply that the perturbed trajectory \mbox{$\mathbf{x}^\prime(t{T^\prime}/{T}) \simeq \mathbf{x}(t) + \mathbf{y}_{\mathrm{P}}(t)\delta p$} obtained from the linear problem may not belong to the attractor of the perturbed system, although it may lie close to it. Hence, statistics computed using $\mathbf{y}_{\mathrm{P}}(t)$ may not be representative of statistics computed on the perturbed attractor, resulting in a sensitivity error that does not vanish as $T\rightarrow\infty$. In our numerical experiments on the Lorenz equations this has materialised as a consistent bias between the periodic shadowing gradients and the finite-difference gradient approximation, as far as the sensitivity with respect to the parameter $\rho$ is concerned. On the other hand, the sensitivity with respect to perturbations of the parameter $\gamma$, equivalent to a smooth coordinate transformation and thus not inducing structural bifurcations, is correctly predicted by linearised methods. A similar breakdown of the method has been observed for the Kuramoto-Sivashinky equation \citep{Blonigan:2014je, Lasagna:2017tz}.

As already argued in Ref.~\citep{Blonigan:2018gd} and references therein, the hope lies in the so called ``chaotic hypothesis'' \citep{GALLAVOTTI:1995kn}. The hypothesis is that structural changes in the attractor as parameters are varied are not that catastrophic if the dimension of the system is large enough, in the ``thermodynamic limit'' \citep{Ruelle:1999bm}. In other words, high-dimensional dynamical systems are ``dynamically stable''; they behave as if they were hyperbolic \cite{Albers:2006df}. However, it currently remains a speculation whether such conjecture can be introduced to provide support to shadowing-based algorithms.  Verifying this hypothesis, and quantifying rigorously how and if such bias varies with the attractor dimension, perhaps by developing scaling laws, would provide great confidence in the application of shadowing based adjoint methods to large scale dynamical systems, e.g. discretisation of spatially extended system such as turbulent flows. Research in this direction is currently in progress and will be reported elsewhere.

\section{Conclusions}\label{eq:conclusions}
The sensitivity of statistical quantities of a hyperbolic chaotic system with respect to parameter perturbations can be determined from the shadowing direction, the unique, uniformly bounded solution of the sensitivity equations. At present, computationally efficient methods to approximate such a direction are hardly available and are essentially limited to variants of the Least Squares Shadowing (LSS) method, initially proposed in Ref.~\citep{Wang:2014hu}.

The major contribution of the current paper is an alternative shadowing-based sensitivity algorithm. Rather than formulating an optimisation problem, as in the LSS method, the heuristic here consists of complementing the linear sensitivity equations with periodic boundary conditions in time. This leads to a boundary value problem in time and requires appropriate numerical methods for the solution, such as the multiple-shooting approach used in this paper.

To provide rigorous support for this new approach, the paper contains a detailed error analysis. We show that, assuming hyperbolicity, our method has the same convergence rate of LSS as the time span $T$ tends to infinity. Specifically, the sensitivity error first decays, on average as $1/T$, and then asymptotically as $1/\sqrt{T}$, the rate at which finite-time averages converge to the infinite-time average. Hence, for $T$ larger then a certain threshold the sensitivity error is dominated by finite-time averaging errors and does not depend on the sensitivity algorithm. We conjecture that these convergence rates are common to all shadowing based algorithms, and thus other considerations, such as accuracy, computational efficiency or robustness to lack of hyperbolicity will come into play.

The theoretical analysis also considers the statistical distribution of the sensitivity error. To the best of our knowledge, this is the first paper that addresses this point for a shadowing-based sensitivity methods. We have shown that, for a given initial condition, the boundary value problem associated to the solution of the sensitivity equations becomes singular on a zero measure set of time spans, i.e.~there exists time spans $T^o_i$, $i=1, \ldots$, for which the sensitivity error is infinite. We show that when this occurs, the least singular value of the associated matrix equation behaves as $k_i|(T-T^o_i)|$, for some positive constant $k_i$. This means that, when a random time span $T$ is used, the probability density function of the sensitivity $\mathcal{J}^T_{\mathrm{d}p}$ displays power-law tails of the form $p(\mathcal{J}^T_{\mathrm{d}p}) \simeq 1/(\mathcal{J}^T_{\mathrm{d}p})^2$. However, we have shown that the probability of large sensitivity errors occurring decrease as $1/T$ for large $T$, so the method converges from a practical point of view.

To support our theoretical predictions, the paper includes numerical calculations on the Lorenz equations for which we have investigated two sensitivity problems. The first involves the sensitivity of statistical quantities with respect to the classical parameter $\rho$. The second arises from defining a non-trivial smooth coordinate transformation, controlled by the parameter $\gamma$. For the latter case, the shadowing direction is known analytically and a full error analysis is possible. Statistically accurate numerical experiments have shown that in the latter case the periodic shadowing sensitivity converges correctly to the value obtained using finite-difference gradient approximations, obtained from long-time averages of chaotic solutions. By contrast, for the sensitivity with respect to $\rho$ we have observed a consistent, reproducible 2\% bias between the finite difference gradient approximation and the sensitivity obtained from all shadowing based algorithms, including periodic shadowing, sensitivity from unstable periodic orbits, and two independent implementations of LSS. As suggested in section \ref{sec:discussion}, this bias is the manifestation of the lack of hyperbolicity.

There are several important aspects requiring further research. For instance, a better understanding of how the proposed method performs in high-dimensional systems is warranted. A better understanding and characterization of the spectral properties of the multiple-shooting system resulting from the periodic shadowing approach is also needed. The spectral characteristics affect the convergence rate of iterative linear algebra solvers \citep{Blonigan:2018gd}, a necessary step forward for high-dimensional PDE discretisations. A further research problem is to quantify if and how sensitivity errors due to lack of hyperbolicity vary with the system's dimension. Finally, alternative strategies to set the gradient $T_{\mathrm{d} p}$ are required, to prevent the boundary value problem (\ref{eq:forward-sensitivity-problem}) and the multiple-shooting system to become singular. This might in turn improve the conditioning of the problem and result in a more favourable probability distribution of the sensitivity. We wish to address these aspects in future work.

\section*{Acknowledgements}
JM acknowledges support of the Flemish Science Foundation (FWO, grant number G0C3115N).

\appendix

\section{Multiple-shooting solution of linear periodic BVPs}\label{sec:multiple-shooting}
We solve the linear periodic BVPs such as (\ref{eq:forward-sensitivity-problem}) and (\ref{eq:adjoint-sensitivity-problem}) using shooting techniques \citep{Ascher:1994ty}. For dynamical systems with unstable dynamics, a multiple-shooting approach is required. These methods are well known \citep{SANCHEZ:2010df,Waugh:2013hq} and we will thus limit the description to the tangent approach.

We define a mesh of $N$ shooting points $t_i = iT/N, i=0, \ldots, N-1$ to partition the time span $[0, T]$ into $N$ segments of equal length.  By linearity, the solution $\mathbf{y}_i(t)$ over the $i$-th segment $t\in[t_i, t_{i+1}]$, originating from a particular initial condition $\mathbf{y}^0_i$, can be written as
\begin{subequations}\label{eq:variation-of-constants-formula}
\begin{empheq}[]{align}
	\mathbf{y}_i(t) =&\; \mathbf{Y}_i(t, \mathbf{x}_i)\cdot\Bigg[\mathbf{y}^0_i + \int_{t_i}^{t_{i+1}} \mathbf{Y}_i^{-1}(s, \mathbf{x}_i)\cdot\Big[\mathbf{f}_{\partial p}(\mathbf{x}(s), p) + T_{\mathrm{d}p}/T\,\mathbf{f}(\mathbf{x}(s), p)\Big]\,\mathrm{d}s \Bigg]\\ =&\; \mathbf{Y}_i(t, \mathbf{x}_i)\cdot\mathbf{y}^0_i + \mathbf{h}_i(t) + (t-t_i)T_{\mathrm{d}p}/T\,\mathbf{f}(\mathbf{x}(t)),\quad  i=0, \ldots, N-1,
\end{empheq}
\end{subequations}
where $\mathbf{x}_i = \mathbf{x}(t_i; \mathbf{x}_0)$ and where $\mathbf{Y}_i(t, \mathbf{x}_i)$ and $\mathbf{h}_i(t)$ are the principal matrix solution and a particular solution, respectively, over the $i$-th segment, solving the initial value problems (IVPs)
\begin{subequations}\label{eq:lpbvp-appendix-fundamental-and-particular}
\begin{empheq}[]{align}
\hspace{1cm}\dot{\mathbf{Y}}_i(t, \mathbf{x}_i) =&\; \mathbf{f}_{\partial \mathbf{x}}(\mathbf{x}(t), p)\cdot\mathbf{Y}_i(t, \mathbf{x}_i),\quad t\in[t_i, t_{i+1}]\; &\mathbf{Y}_i(t_i, \mathbf{x}_i) =&\; \mathbf{I}\hspace{0.7cm}\\
\dot{\mathbf{h}_i}(t) =&\; \mathbf{f}_{\partial \mathbf{x}}(\mathbf{x}(t), p)\cdot \mathbf{h}_i(t) + \mathbf{f}_{\partial p}(\mathbf{x}(t), p),\quad t\in[t_i, t_{i+1}]\; &\mathbf{h}_i(t_i) =&\; \mathbf{0},\label{eq:lpbvp-appendix-fundamental-and-particular-part}
\end{empheq}
\end{subequations}
for $i=0, \ldots, N-1$ and where $\mathbf{I}$ and $\mathbf{0}$ are the identity matrix and a vector of zeros of appropriate size. The principal matrix solution propagates vectors in the tangent space forward in time, i.e.~for any vector satisfying the variational equations
\begin{equation}\label{eq:variational-equations}
	\dot{\mathbf{v}}(t) = \mathbf{f}_{\partial \mathbf{x}}(\mathbf{x}(t), p)\cdot \mathbf{v}(t), \quad t \in [t_i, t_{i+1}],\quad \mathbf{v}(0) = \mathbf{v}^{0},
\end{equation}
for some $i$, $\mathbf{v}(t) = \mathbf{Y}_i(t, \mathbf{x}_i)\cdot\mathbf{v}^{0}$ holds. Since the matrices $\mathbf{Y}_i(t, \mathbf{x}_i)$, $t\geq t_i$ are invertible for all $t\geq t_i$ \citep{hale2009ordinary}, their inverses map tangent vectors backwards in time, i.e.~$\mathbf{v}^{0} = \mathbf{Y}_i^{-1}(t, \mathbf{x}_i)\cdot\mathbf{v}(t)$ holds, $\forall t\geq t_i$. This allows the term with $\mathbf{f}(\mathbf{x}(t))$ to be integrated exactly, since $\mathbf{Y}_i^{-1}(s, \mathbf{x}_i)\cdot\mathbf{f}(\mathbf{x}(s), p) = \mathbf{f}(\mathbf{x}_i, p)$, $s \in [t_i, t_{i+1}]$.

We then seek the initial conditions $\mathbf{y}^0_i, i=0, \ldots, N-1$, such that the overall solution is continuous at the shooting points. Using the general form of the solution (\ref{eq:variation-of-constants-formula}) and imposing the continuity conditions 
\begin{equation}
	\mathbf{y}^0_{i+1} = \mathbf{Y}_i(t_{i+1}, \mathbf{x}_i)\cdot\mathbf{y}^0_i + \mathbf{h}_i(t_{i+1}) + (t_{i+1}-t_i)T_{\mathrm{d}p}/T\,\mathbf{f}(\mathbf{x}_{i+1}, p),\quad  i=0, \ldots, N-1,\\
\end{equation}
with suitable modification for the case $i=N-1$, leads to the bordered system of linear equations
\begin{equation}\label{eq:linear-system-multiple-shooting}
	\left[\begin{array}{cccc|c}
	\phantom{-}\mathbf{Y}_0(t_1, \mathbf{x}_0)  & - \mathbf{I}               &        &                                  & \mathbf{f}(t_1) \\[4pt]
	                              &    \ddots                  & \ddots &                                  & \vdots\\[4pt]
	                              &                            & \ddots &  - \mathbf{I}                    & \vdots\\[4pt]
	-\mathbf{I}                   &                            &        & \phantom{-}\mathbf{Y}_{N-1}(t_N, \mathbf{x}_{N-1}) & \mathbf{f}(t_N) \\[4pt]\hline\!\rule{0in}{.5cm}
	\phantom{-}\mathbf{f}^\top(0) & \phantom{-}\mathbf{0}^\top &\ldots  & \phantom{-}\mathbf{0}^\top       & 0          \\
	\end{array} \right]\cdot
	\left[\begin{array}{c}
	\mathbf{y}^0_0 \\[4pt]
	\vdots\\[4pt]
	\vdots\\[4pt]
	\mathbf{y}^0_{N-1} \\[4pt]\hline\!\rule{0in}{.5cm} 
	T_{\mathrm{d} p}^0/N
	\end{array}\right] = 
	-\left[\begin{array}{c}
	\mathbf{h}_0(t_1) \\[4pt]
	\vdots\\[4pt]
	\vdots\\[4pt]
	\mathbf{h}_{N-1}(t_{N}) \\[4pt]\hline\!\rule{0in}{.5cm}
	 0
	\end{array}\right],
\end{equation}
where the bordering vectors are included to enforce the orthogonality constraint (\ref{eq:forward-sensitivity-problem-c}). 

The number of shooting stages, i.e.~the quantity $T/N$, controls the condition number of the matrix at the left-hand-side of (\ref{eq:linear-system-multiple-shooting}) and the accuracy of the numerical solution. For the Lorenz problem discussed in section \ref{sec:numerical-example}, we typically used a constant $T/N$ no greater than 5 time units, to to satisfy the condition $|\mathbf{y}(0)-\mathbf{y}(T)|/|\mathbf{y}(0)| < 10^{-8}$, with $dt=10^{-3}$.

For low-dimensional systems, such as in the present case, the principal matrix solutions $\mathbf{Y}_{i}(t_{i+1}, \mathbf{x}_i)$ can be constructed by solving the $N$ IVPs associated to (\ref{eq:lpbvp-appendix-fundamental-and-particular}) using the canonical basis vectors of $\mathbb{R}^N$ as initial conditions. Efficient dense linear algebra techniques that leverage the bordered, banded structure of the left-hand-side of (\ref{eq:linear-system-multiple-shooting}) can then be used (see \citep{Ascher:1994ty, Govaerts:2000vv}). However, for the Lorenz system considered later in the present work, we simply employed a standard LU factorisation technique. 

For discretisation of PDEs, the construction and factorisation of the matrix in the left-hand-side can quickly become prohibitive. Iterative Krylov subspace methods that do not require the matrix to be constructed but only its action on a vector, should instead be used. In such case, the action of the operators $\mathbf{Y}_i(t_{i+1}, \mathbf{x}_i)$ on the elements $\mathbf{y}^0_i$ can be computed using a tangent code, preferably simultaneously over the $N$ segments, in a time-parallel fashion. The formulation of appropriate algorithms for PDEs is currently in progress and will be reported in future publications. 

\section{Calculation of $T^0_{\mathrm{d} p}$ in the adjoint method}\label{app:adjoint-tp0}
For the adjoint method, the gradient $T^0_{\mathrm{d} p}$ is obtained from a solvability condition for the problem (\ref{eq:forward-sensitivity-problem}) akin to the well known Fredholm's alternative for the solution of linear systems. Specialising the general solution of (\ref{eq:forward-sensitivity-problem-a}),
\begin{equation}
	\mathbf{y}(t) = \mathbf{Y}(t, \mathbf{x}_0)\cdot\bigg[\mathbf{y}(0) + \int_{0}^{t} \mathbf{Y}^{-1}(s, \mathbf{x}_0)\cdot\mathbf{f}_{\partial p}(\mathbf{x}(s), p)\mathrm{d}s\bigg] + t\frac{T_{\mathrm{d}p}}{T}\mathbf{f}(\mathbf{x}(t), p),
\end{equation}
at $t=T$ and using the boundary condition (\ref{eq:forward-sensitivity-problem-b}) and the orthogonality constraint (\ref{eq:forward-sensitivity-problem-c}), the BVP (\ref{eq:forward-sensitivity-problem}) can be formally transformed into the problem
\begin{subequations}\label{eq:period-gradient-adjoint-solve-st}
\begin{empheq}[]{align}
\mathrm{solve}:\quad [\mathbf{Y}(T, \mathbf{x}_0) - \mathbf{I}]\cdot\mathbf{y}(0) &= -\mathbf{h}(T) - T^0_{\mathrm{d} p}\mathbf{f}(T)\label{eq:period-gradient-adjoint-solve-st-equation}\\
\mathrm{subject\;to}:\hspace{1.5cm} \mathbf{f}(0)^\top\cdot\mathbf{y}(0)&=0\label{eq:period-gradient-adjoint-solve-st-constraint}
\end{empheq}
\end{subequations}
where 
\begin{equation}
	\mathbf{h}(T) = \mathbf{Y}(T, \mathbf{x}_0)\cdot\int_{0}^{T} \mathbf{Y}^{-1}(s, \mathbf{x}_0)\cdot\mathbf{f}_{\partial p}(\mathbf{x}(s), p)\mathrm{d}s.
\end{equation}
Assuming invertibility of $\mathbf{Y}(T, \mathbf{x}_0)-\mathbf{I}$, the solution of (\ref{eq:period-gradient-adjoint-solve-st-equation}) can be substituted in the constraint (\ref{eq:period-gradient-adjoint-solve-st-constraint}) to obtain
\begin{equation}
	\Big[ [\mathbf{Y}(T, \mathbf{x}_0) - \mathbf{I}]^{-1}\cdot \big[\mathbf{h}(T) + T^0_{\mathrm{d} p}\mathbf{f}(T)\big]\Big]^{\top}\cdot\mathbf{f}(0) = 0.
\end{equation}
This can be further rearranged into
\begin{equation}\label{eq:homogenous-adjoint-problem-equality}
	\big[\mathbf{h}(T) + T^0_{\mathrm{d} p}\mathbf{f}(T)\big]^\top\cdot\widecheck{\boldsymbol \lambda}(T) =0
\end{equation}
where the vector $\widecheck{\boldsymbol \lambda}(T)\in\mathbb{R}^N$ is the solution of the adjoint matrix equation
\begin{equation}\label{eq:homogeneous-adjoint-problem-matrix-equation}
	[\mathbf{Y}(T, \mathbf{x}_0)^\top - \mathbf{I}\big]\cdot\widecheck{\boldsymbol \lambda}(T) = \mathbf{f}(0),
\end{equation}
which corresponds in practice to solving the following homogeneous adjoint problem 
\begin{equation}\label{eq:homogeneous-adjoint-problem}
	\dot{\widecheck{\boldsymbol \lambda}}(t) = -\mathbf{f}^\top_{\partial\mathbf{x}}(\mathbf{x}(t))\cdot \widecheck{\boldsymbol \lambda}(t), \quad t\in[0, T], \quad \widecheck{\boldsymbol \lambda}(T) = \widecheck{\boldsymbol \lambda}(0) + \mathbf{f}(0).
\end{equation}
For numerical purposes, it may be convenient to transform (\ref{eq:homogeneous-adjoint-problem}) into a problem with periodic boundary conditions. This can be done by writing the solution as \mbox{$\widecheck{\boldsymbol \lambda}(t) = {\boldsymbol q}(t) - t/T\,\mathbf{f}(0)$}, where ${\boldsymbol q}(t)$ satisfies ${\boldsymbol q}(0) = {\boldsymbol q}(T)$, and then solving the standard problem
\begin{subequations}\label{eq:adjoint-sensitivity-problem-tp}
\begin{empheq}[left={\empheqlbrace}]{alignat=1}
\dot{\boldsymbol q}(t) &= -\mathbf{f}_{\partial \mathbf{x}}^\top(\mathbf{x}(t), p)\cdot{\boldsymbol q}(t) + \frac{1}{T}\big(\mathbf{f}(0) + t\, \mathbf{f}_{\partial \mathbf{x}}^\top(0)\cdot\mathbf{f}(0)\big), \quad t \in [0, T], \\
{\boldsymbol q}(0) &={\boldsymbol q}(T). 
\end{empheq}
\end{subequations}
The gradient $T_{\mathrm{d}p}^0/T$ can then be obtained from (\ref{eq:homogenous-adjoint-problem-equality}) with the quadrature
\begin{equation}\label{eq:period-gradient-formula}
	\displaystyle \frac{T^0_{\mathrm{d} p}}{T} = \displaystyle -\frac{\displaystyle \frac{1}{T}\int_0^T \widecheck{\boldsymbol \lambda}(t)^\top\cdot \mathbf{f}_{\partial p}(t)\,\mathrm{d}t}{\widecheck{\boldsymbol \lambda}(T)^\top\cdot \mathbf{f}(T)}
\end{equation}
where the fact that $\mathbf{Y}(T, \mathbf{x}_0)^\top\cdot\widecheck{\boldsymbol \lambda}(T) = \widecheck{\boldsymbol \lambda}(0)$ and $\mathbf{Y}(s, \mathbf{x}_0)^{-\top}\cdot\widecheck{\boldsymbol \lambda}(0) = \widecheck{\boldsymbol \lambda}(s)$ is used to simplify the numerator of (\ref{eq:period-gradient-formula}).

In short, the adjoint periodic shadowing method requires the solution of an additional adjoint problem in addition to the main adjoint problem, equation (\ref{eq:adjoint-sensitivity-problem}). We expect, however, the overall computational cost of solving the two problems to be somewhere between one and two times the cost of solving one adjoint problem, depending on the numerical method used and the sophistication of thee implementation. For low-dimensional dynamical systems, where the multiple-shooting matrix is first constructed and then factorised, the major source of cost is the repeated integration of the linearised equations to construct the principal matrix solutions associated to the adjoint equations. The LU factorisation of this matrix can be then reused to solve the two linear problems at essentially no cost. For high-dimensional systems, e.g. discretisations of partial differential equations, where iterative techniques will be required for solving the multiple-shooting system, an efficient implementation would integrate the two adjoint problems jointly, reusing as much as possible auxiliary calculations involved in the construction of the adjoint of the linearised operator. 

\section{Singularity conditions of the bordered system}\label{app:B-singular}
In this appendix we show that the bordered matrix $\mathbf{B}(T, \mathbf{x}_0)$ in equation (\ref{eq:error-bordered-system}) becomes singular when
\begin{equation}
\mathbf{f}(0)^\top\cdot\mathbf{M}^{-1}(T, \mathbf{x}_0)\cdot\mathbf{f}(T)=0.	
\end{equation}
Singularity occurs when one can find a non trivial vector $\hat{\mathbf{w}} = [\mathbf{w}^\top, w]^\top \in \mathbb{R}^{N+1}$, $w \in \mathbb{R}$, such that $\mathbf{B}(T, \mathbf{x}_0)\cdot\hat{\mathbf{w}} = \mathbf{0}$. Expanding the two block rows of $\mathbf{B}(T, \mathbf{x}_0)$, this is equivalent to finding $\mathbf{w}$ and $w$ such that
\begin{align}
	\mathbf{M}(T, \mathbf{x}_0)\cdot{\mathbf{w}} + w\mathbf{f}(T) =\;& \mathbf{0}\label{eq:B-singular-1},\\
	\mathbf{f}(0)^\top\cdot\mathbf{w} =\;& 0\label{eq:B-singular-2}.
\end{align}
It can be shown that $\mathbf{M}(T, \mathbf{x}_0)$ is generally invertible when $\mathbf{B}(T, \mathbf{x}_0)$ is singular. Hence, the first condition implies that
\begin{equation}
	{\mathbf{w}} = - w\mathbf{M}^{-1}(T, \mathbf{x}_0)\cdot\mathbf{f}(T).
\end{equation}
Taking the dot product of this expression with $\mathbf{f}(0)$ and using (\ref{eq:B-singular-2}) leads to
\begin{equation}
	w\mathbf{f}(0)^\top\cdot\mathbf{M}^{-1}(T, \mathbf{x}_0)\cdot\mathbf{f}(T) = 0,
\end{equation}
which holds for any non trivial $w$ if and only $\mathbf{f}(0)^\top\cdot\mathbf{M}^{-1}(T, \mathbf{x}_0)\cdot\mathbf{f}(T)=0$.

\section{Behaviour of the least singular value of the bordered system}\label{app:svd-near-a-zero}
We will use the Singular Value Decomposition (SVD)
\begin{equation}
	\mathbf{B}(T, \mathbf{x}_0) = \left[\begin{array}{c|c}\mathbf{U}&\mathbf{u}_0\end{array}\right]\left[\begin{array}{c|c}\mathbf{\Sigma}&0\\[1pt]
	\hline\!\rule{0in}{.4cm}0&\sigma_0\end{array}\right]\left[\begin{array}{c|c}\mathbf{V}&\mathbf{v}_0\end{array}\right]^\top,
\end{equation}
where we omit for clarity the dependence of singular values and singular vectors from $T$ and $\mathbf{x}_0$.
For some initial condition $\mathbf{x}_0$, let now consider a time span $T^o$ for which the matrix $\mathbf{B}(T^o, \mathbf{x}_0)$ is singular. In these conditions, the first right singular vector $\mathbf{v}_0$ solves the optimisation problem
\begin{equation}\label{eq:svd-optimisation}
	\sigma_{0}^2 = \min_{\mathbf{v}_0} \|\mathbf{B}(T^o, \mathbf{x}_0)\cdot\mathbf{v}_0\|^2 = 0,\quad\mathrm{subject~to}\quad\|\mathbf{v}_{0}\| = 1.
\end{equation}
At first order, a small perturbation of the time span $\delta T$ produces a small perturbation $\delta\mathbf{B}$, resulting in a perturbation of the singular vector $\delta\mathbf{v}_0 $ and in a perturbation of the least singular value $\delta\sigma_0$. To find such a perturbation, we would solve the problem
\begin{equation}\label{eq:svd-optimisation-pert}
	(\sigma_{0} + \delta  \sigma_0)^2 = \delta\sigma_0^2 = \min_{\delta \mathbf{v}_0} \|(\mathbf{B}(T^o, \mathbf{x}_0) + \delta \mathbf{B})\cdot(\mathbf{v}_0 + \delta\mathbf{v}_0)\|^2,\quad\mathrm{subject~to}\quad\delta\mathbf{v}_{0}^\top\cdot\mathbf{v}_0 = 0,
\end{equation}
where the orthogonality arises from the fact that a small perturbation $\delta T$ induces a small rotation of the orthogonal basis formed by the left singular vectors. At first order, $\delta \mathbf{v}_0 = \tilde{\mathbf{v}}_0\delta T$ and 
\begin{equation}
	\delta\mathbf{B} = \left[\begin{array}{c|c} \delta{\mathbf{M}} & \delta{\mathbf{f}}(T^o) \\[1pt]
	\hline\!\rule{0in}{.4cm}0 & 0 \end{array}\right] = \left[\begin{array}{c|c} \tilde{\mathbf{M}} & \tilde{\mathbf{f}} \\[1pt]
	\hline\!\rule{0in}{.4cm}0 & 0 \end{array}\right]\delta T = \tilde{\mathbf{B}}\delta T,
\end{equation}
where it can be shown that $\tilde{\mathbf{M}} = \mathbf{Y}(T^o, \mathbf{x}_0)\cdot\mathbf{f}_{\partial\mathbf{x}}(T^o)$ and $\tilde{\mathbf{f}} = \mathbf{f}_{\partial\mathbf{x}}(T^o)\cdot\mathbf{f}(T)$. Expanding now the product in (\ref{eq:svd-optimisation-pert}), and neglecting higher order terms, the optimisation problem becomes 
\begin{equation}
	\delta\sigma_0^2 = \min_{\tilde{\mathbf{v}}_0} \| \mathbf{B}(T^o, \mathbf{x}_0)\cdot\tilde{\mathbf{v}_0} + \tilde{\mathbf{B}}\cdot\mathbf{v}_0\|^2\delta T^2,\quad\mathrm{subject~to}\quad \tilde{\mathbf{v}}_0^\top\cdot\mathbf{v}_0 = 0.
\end{equation}
Since the term in the norm does not depend on $\delta T$, we conclude that near a singular point $T^o$, the least singular value behaves as $|k(T-T^o)|$, where the constant $k$ is 
\begin{equation}\label{eq:optimisation-k}
	k^2 = \min_{\tilde{\mathbf{v}}_0} \| \mathbf{B}(T^o, \mathbf{x}_0)\cdot\tilde{\mathbf{v}}_0 + \tilde{\mathbf{B}}\cdot\mathbf{v}_0\|^2,\quad\mathrm{subject~to}\quad \tilde{\mathbf{v}}_0^\top\cdot\mathbf{v}_0 = 0.
\end{equation}
Expanding $\tilde{\mathbf{v}}_0$ into the right singular vectors $\mathbf{V}$ to lift the orthogonality constraint in (\ref{eq:optimisation-k}) and substituting back, it can be shown that the minimum of this optimisation problem is
\begin{equation}
	k = \mathbf{u}_0^\top\cdot\tilde{\mathbf{B}}\cdot\mathbf{v}_0,
\end{equation}
i.e.~the error resulting from projecting the vector $\tilde{\mathbf{B}}\cdot\mathbf{v}_0$ onto the left singular vectors $\mathbf{U}$. Note that the constant $k$ is different from zero almost always, since there is in general no particular relation between the SVD of $\mathbf{B}(T, \mathbf{x}_0)$ and the perturbation $\tilde{\mathbf{B}}$. 

The same procedure outlined here can be used to show that the least singular value of the matrix $\mathbf{M}(T, \mathbf{x}_0)$ displays the same type of zeros as the bordered system.

\section{Deriving the power-law tail for $1/\sigma_0(T, \mathbf{x}_0)$}\label{app:power-law}
We show that sampling a positive function $g(x) : \mathbb{R}\rightarrow\mathbb{R}$ possessing multiple zeros like $g(x) \simeq k_i|(x - x^o_i)|$ near $x^o_i$, $i = 1,\ldots$, leads to a probability density function for $1/g(x)$ with a power-law right tail of the form $p(1/g(x)) \simeq 1/g(x)^2$. 

The probability that $g(x)$ is less than some constant $R$ can be expressed using the cumulative probability density function $P(g(x) < R)$. For small $r$, the linearity of $g(x)$ near the zeros implies that
\begin{equation}
	P(g(x) < r) \simeq cr,\quad r \ll 1
\end{equation}
for some constant $c > 0$ that might depend on the frequency of the zeros and the average slope of $g(x)$ near them. With the change of variable $R=1/r$, we obtain
\begin{equation}
	P(g(x) < 1/R) \simeq {c}/{R}, \quad r \gg R.
\end{equation}
Now, since $P(g(x) < 1/R) = P(1/g(x) > R) = 1 - P(1/g(x) < R)$, we obtain that
\begin{equation}\label{eq:cumulative}
	P(1/g(x) < R) \simeq 1 - {c}/{R}, \quad R \gg 1
\end{equation}
The asymptotic behaviour of the probability density function $p(1/g(x))$ for small $g(x)$ is now readily obtained by differentiation of the cumulative distribution (\ref{eq:cumulative}) as
\begin{equation}
	p(1/g(x)) \simeq c/{g(x)^2}, \quad g(x) \ll 1.
\end{equation}

\section{Analysis of the bordered system for a chaotic aero-elastic oscillator}\label{sec:aero-elastic}
This last appendix considers the spectral properties of the matrix $\mathbf{B}(T, \mathbf{x}_0)$ for the aero-elastic oscillator previously used as a test bed for shadowing methods \citep{Wang:2014hu, Liao:2016hn}.
\begin{figure}[tbp]
	\centering
	\includegraphics[width=0.95\textwidth]{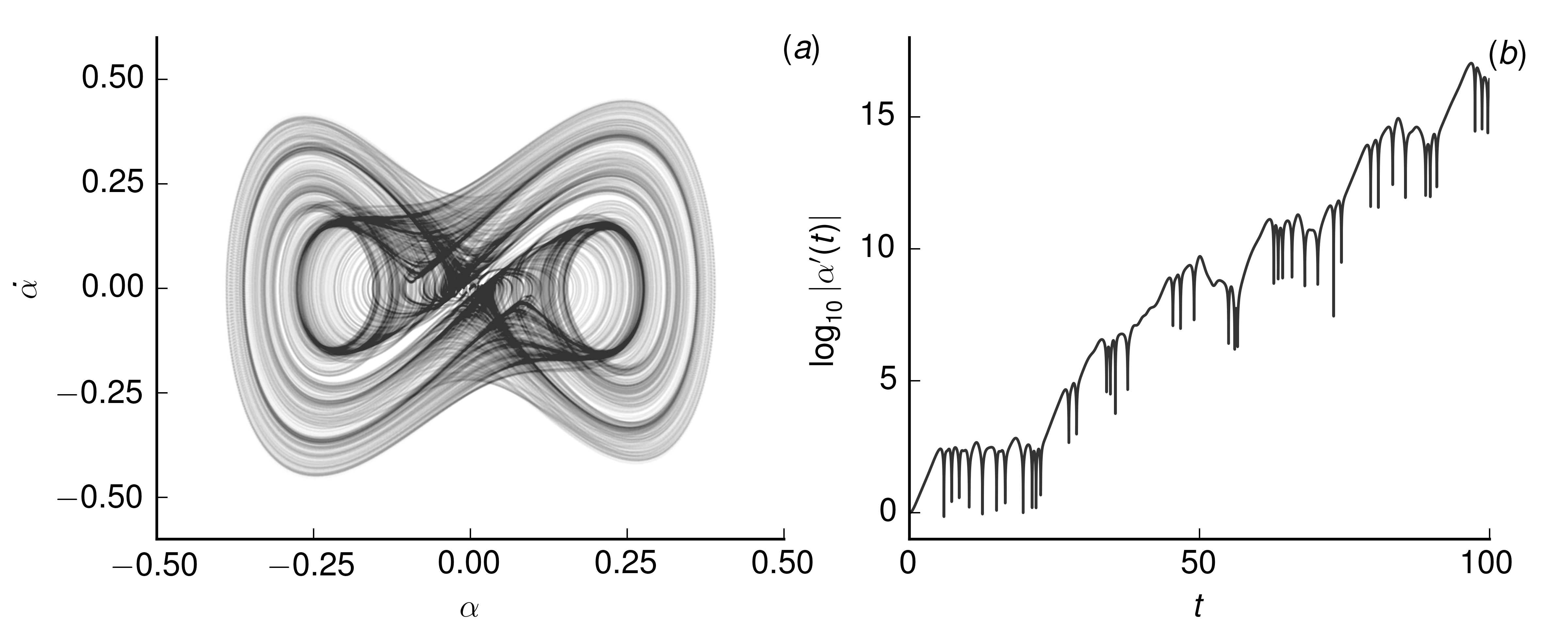}
	\caption{State-space trajectory of the aero-elastic oscillator of equation (\ref{eq:aero-equation}), panel (a), and growth of the pitch perturbation from the initial condition $\mathbf{y}_0 = [0, 1, 0, 0]^\top$, for a point $\mathbf{x}_0$ on the attractor, panel (b).}
	\label{fig:aero-oscill-attractor}
\end{figure}
The dynamics of the oscillator are defined by the second order nonlinear differential equation
\begin{equation}\label{eq:aero-equation}
\mathbf{G}\cdot\ddot{\mathbf{z}}(t) + \mathbf{D}\cdot\dot{\mathbf{z}}(t)+ (\mathbf{K}_1 + \mathbf{K}_2 Q)\cdot\mathbf{z}(t) + \mathbf{K}_3\cdot\mathbf{z}^3(t) = 0,
\end{equation}
where the state $\mathbf{z}(t) = [h(t),\;\alpha(t)]^\top$ is defined by the plunge and pitch degrees of freedom and  
\begin{equation}
\mathbf{G} = \frac{1}{4}\begin{bmatrix}4 &1 \\ 1 &2\end{bmatrix},\;\;\mathbf{D} = \frac{1}{10}\begin{bmatrix}1 &0 \\ 0 &1\end{bmatrix},\;\;\mathbf{K}_1 = \frac{1}{10}\begin{bmatrix}2 &0 \\ 0 &5\end{bmatrix},\;\;\mathbf{K}_2 = \frac{1}{10}\begin{bmatrix}0 &1 \\ 0 &-1\end{bmatrix},\;\;\mathbf{K}_3 = \begin{bmatrix}0 &0 \\ 0 &20\end{bmatrix},
\end{equation}
while $Q$ is the bifurcation parameter. The second order equation (\ref{eq:aero-equation}) is transformed into a set of four first order equations, by defining the state vector $\mathbf{x}(t) = [h(t), \alpha(t), \dot{h}(t), \dot{\alpha}(t)]^\top$, with the associated perturbation vector $\mathbf{y}(t) = [h^\prime(t), \alpha^\prime(t), \dot{h}^\prime(t), \dot{\alpha}^\prime(t)]^\top$. Here, we consider $Q=11.2$, at which chaotic long-term behaviour is observed, as illustrated in figure \ref{fig:aero-oscill-attractor}.

\begin{figure}[htbp]
	\centering
	\includegraphics[width=0.99\textwidth]{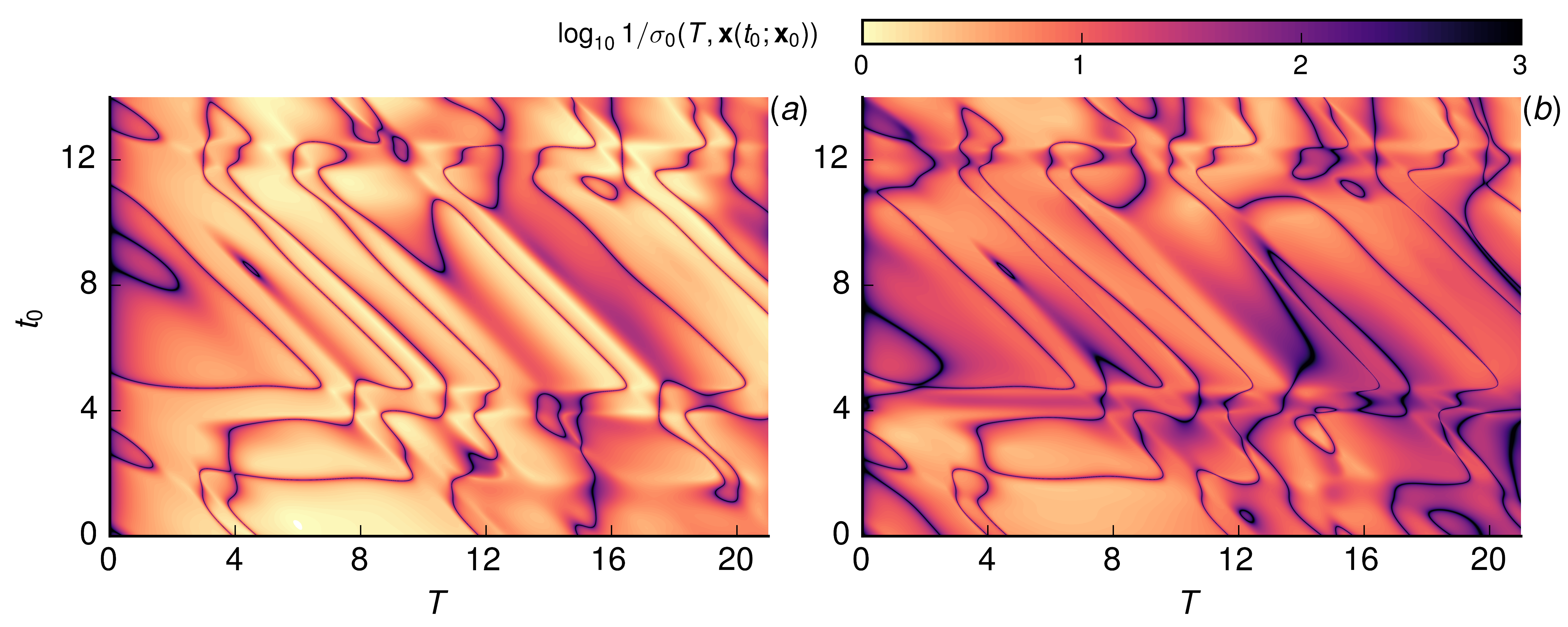}
	\caption{Distribution of the inverse of the least singular value for the matrix $\mathbf{M}(T, \mathbf{x}(t_0; \mathbf{x}_0))$, (a), and of the bordered matrix $\mathbf{B}(T, \mathbf{x}(t_0; \mathbf{x}_0))$, (b), for the aero-elastic oscillator problem. The least singular value can be smaller than $10^{-3}$, but the colour map has been clipped at this value for better clarity.}
	\label{fig:aero-elastic-svd} 
\end{figure}

We first obtain a point $\mathbf{x}_0$ on the attractor by integrating the governing equations (same settings as for the Lorenz equations) for a sufficiently long time for transients to decay. We then construct the matrices $\mathbf{M}(T, \mathbf{x}(t_0; \mathbf{x}_0))$ and $\mathbf{B}(T, \mathbf{x}(t_0; \mathbf{x}_0))$ for a range of $T$ and $t_0$, as discussed in section \ref{eq:singularity-conditions},  and calculate the least singular value, denoted as $\sigma_0(T, \mathbf{x}(t_0; \mathbf{x}_0))$. Results are reported in figure \ref{fig:aero-elastic-svd} for the non-bordered system, panel (a), and for the bordered system, panel (b). As in figure (\ref{fig:singular-values-of-bordered-nonbordered-systems})-(a), we plot the base ten logarithm of the inverse of $\sigma_0$ to better highlight singularity conditions. Similarly to the Lorenz equations, the matrix $\mathbf{M}$ becomes singular on a zero measure set of time spans for a given $t_0$. Although not shown here, this occurs precisely when condition (\ref{eq:chi-definition}) holds. The major difference with the Lorenz equations, though, is that the bordered system (\ref{eq:error-bordered-system}) becomes singular on a zero measure set of pairs $(T, t_0)$, precisely where (\ref{eq:chi-definition}) holds. It is argued that the bordered system becomes occasionally singular for many chaotic dynamical systems, with the Lorenz equations being an exception. Although not reported here for the sake of brevity, we have performed the same analysis on the Kuramoto-Sivashinky equation, a well known one-dimensional PDE with chaotic solutions, using the same setup used in Ref.~\citep{Lasagna:2017tz}. For this problem, we have observed the same structure of figure \ref{fig:aero-elastic-svd}-(b).

\bibliographystyle{plain}
\bibliography{library}

\end{document}